\documentclass[aps,prd,reprint,superscriptaddress,showpacs,nofootinbib]{revtex4-1} % use \documentstyle for old LaTeX compilers

\usepackage[english]{babel}
\usepackage{amssymb}
\usepackage{multirow}
\usepackage{amsmath}
\usepackage{txfonts}
\usepackage{mathdots}
\usepackage[classicReIm]{kpfonts}
\usepackage{graphicx}
\usepackage{float}
\usepackage[utf8]{inputenc}
\usepackage[colorlinks=true]{hyperref}
\usepackage{array}
\usepackage{todonotes}
\usepackage[letterpaper,top=2cm,bottom=2cm,left=3cm,right=3cm,marginparwidth=1.75cm]{geometry}
\usepackage{caption}
\usepackage{subcaption}

\newcommand{\eq}[2]{\begin{equation} \label{eq:#1} #2 \end{equation}}
\newcommand{\meq}[2]{\begin{multline}\label{eq:#1} #2 \end{multline}}
\newcommand{\seq}[2]{\eq{#1}{\begin{split} #2 \end{split}}}
\newcommand{\Eref}[1]{(\ref{eq:#1})}

\begin{document}

\title{Asymptotically flat vacuum solution for a rotating black hole in a modified 
gravity theory}
%\date{}
\author{Arghya Ranjan Das}
\email{arghyadas@iisc.ac.in}
\author{Banibrata Mukhopadhyay}
\email{bm@iisc.ac.in}
\affiliation{Department of Physics, Indian Institute of Science, Bangalore 
560012}

\begin{abstract}
The theory of $f(R)$-gravity is one of the theories of modified Einstein gravity.
The vacuum solution, on the other hand, of the field equation is the solution for black hole
geometry. We establish here an asymptotically flat rotating black hole solution in
an $f(R)$-gravity. This essentially leads to the modified solution to the Kerr black hole.
This solution exhibits the change in fundamental properties of the black hole and its
geometry. It particularly shows that radii of marginally stable and bound orbits and black hole
event horizon increase compared to those in Einstein gravity, depending on the
modified gravity parameter. It further argues for faster spinning black holes with spin
(Kerr) parameter greater than unity, without any naked singularity.
This supports the weak cosmic censorship hypothesis.
\end{abstract}

\maketitle

\section{Introduction}

General relativistic gravity of Einstein turns out to be a remarkable discovery to explain 
a range of astrophysical sources, apart from its theoretical integrity, even after more
than 100 years of its original discovery. Eventually, all the predictions of Einstein's
gravity proved to be correct, particularly after the direct detection of gravitational wave
in 2015 \cite{GW15}. In fact, the said discovery could be considered `three in one': 
direct confirmation of gravitational wave, spinning black hole and binary black hole.

Although to understand coalescence of, e.g., black holes and to probe the underlying gravitational
radiation, strong field general relativity (GR) or numerical relativity is indispensable, most of 
the direct tests of GR are done based on weak field approximation. Therefore,
the global validity of GR in the strong field regime, i.e. the true nature of gravity close
to the source of gravity, remains questionable. Hence, no one can rule out possible modification 
to GR in natural systems, particularly when the theory is asymptotically flat. Asymptotic flatness
assures reduction of modified GR to GR and to Minkowskian with distance from the source. 
Therefore, even if close to the source, i.e. a compact object like black hole, neutron star, actual
gravitational theory is modified GR, the same theory will be able to explain any solar-system based or
Earth-based experiment.

One such example of modified GR is the theory of f(R)-gravity \cite{Upasana15,Kalita18}, which was explored to explain sub- and super-Chandrasekhar limiting mass
white dwarfs in a unified theory,
what GR as such could not.
They are possibly leading to under- and over-luminous type Ia supernovae under the same model
framework. Recently, we also established an asymptotically flat vacuum solution, unlike that for
a white dwarf, of $f(R)$-gravity in spherical symmetry \cite{Kalita_Bani}. This is essentially 
a modified solution for the Schwarzschild, hence nonrotating, black hole. We showed that depending
on the modified gravity parameter, various basic characteristics of the black hole, e.g. marginally
stable and bound circular orbits, event horizon etc., change. We also showed that for a very hot 
accretion flow, critical/sonic point location changes in modified GR. There are other explorations
of black hole in modified GR as well \cite{Nojiri:2013su,Nojiri:2017kex,Nojiri:2020blr}.

However, most of the cosmic objects are rotating, hence more realistic, at least in general, 
black holes are expected to be rotating. The same goes with other compact objects described by
non-vacuum solutions. What if, a black hole is rotating in modified GR, more precisely in
$f(R)$-gravity? In other words, how the Kerr solution changes in the $f(R)$-gravity?

In this work, we establish an asymptotically flat solution for a rotating black hole in modified GR.
In place of obtaining a solution from the appropriate Einstein action for a modified GR, we rely on
the Newman-Janis algorithm (NJA) \cite{NJA_Orig}. We know that based on NJA the Kerr
black hole solution can be derived from the Schwarzschild solution by making an elementary transformation involved with complex numbers. The basic idea is, as if due to the choice of coordinates combining realistic
coordinates and metric parameters, the Kerr metric appeared
to be diagonal and also spherical symmetric, like the Schwarzschild black hole. However, once it is expanded 
in realistic coordinates it turns out to have off-diagonal terms with axially symmetric nature of
the metric. We plan to implement NJA in the modified Schwarzschild metric under $f(R)$-gravity
\cite{Kalita_Bani} to obtain the corresponding modified Kerr solution. To the best of our knowledge, there is no
venture towards this solution before this work. Once we obtain the modified Kerr solution,
we explore various basic characteristics of the metric, e.g. radius of event horizon, 
marginally stable and bound circular orbits, various components of epicyclic oscillation
frequency, orbital angular frequency, etc., with the change of black hole spin and modified
gravity parameter.

The paper is organized as follows. In the next two sections, we recapitulate the basic formalism
of obtaining modified GR based field equation in $f(R)$-gravity and its solution for an asymptotically flat non-rotating
black hole, respectively, in sections II and III. Thereafter, we establish a rotating black hole 
solution in section IV based on NJA. Further, we discuss the nature of singularity of the metric
and horizons in, respectively, sections V and VI. For the latter, first we present the numerical
solution and then approximate analytical solution. Subsequently, we explore various fundamental
orbits, as in GR, in this modified gravity framework for a test particle motion in section VII and
corresponding fundamental oscillation frequencies in section VIII. We conclude our work in section IX.

%%%%%%%%%%%%%%%%%%%%%%%%%%%%%%%%%%%%%%%%%%%%%%%%%%%
%%%%%%%%%%%%%%%%%%%%%%%%%%%%%%%%%%%%%%%%%%%%%%%%%%%
%%%%%%%%%%%%%%%%  Basic Formalism  %%%%%%%%%%%%%%%%
%%%%%%%%%%%%%%%%%%%%%%%%%%%%%%%%%%%%%%%%%%%%%%%%%%%
%%%%%%%%%%%%%%%%%%%%%%%%%%%%%%%%%%%%%%%%%%%%%%%%%%%
\section{Basic formalism of field equation}
In GR, the Einstein-Hilbert action produces the field equation. With the metric signature $(+---)$ in 4-dimension it is given by \cite{Misner_Thorne_Wheeler_73}

\eq{EH_action}{S = \int \left[ \frac{c^4}{16\pi G}R + \mathcal{L}_M\right], }

\noindent where $c$ is the speed of light, $R$ is the scalar curvature 
such that $R=R_{\mu\nu}g^{\mu\nu}$, often 
called Ricci scalar, with $R_{\mu\nu}$ being Ricci tensor, $G$ is Newton's gravitation constant, $\mathcal{L}_M$ is the Lagrangian of the matter field and $g = det(g_{\mu\nu})$ is the determinant of the metric tensor $g_{\mu\nu}$. Varying this action w.r.t. $g_{\mu\nu}$ and equating it to zero with appropriate boundary condition produces the Einstein’s field equation for GR, given by

\eq{Field_eq}{G_{\mu\nu} = R_{\mu\nu}-\frac{1}{2}g_{\mu\nu}R = \frac{8\pi G}{c^4}T_{\mu\nu},}

\noindent where $T_{\mu\nu}$ is the energy-momentum tensor of the matter field. This equation relates the matter to the curvature of the spacetime.
In case of modified GR, here $f(R)$ gravity, the Ricci scalar in Einstein-Hilbert action is replaced by $f(R)$ (being a function of the Ricci scalar). The action is then represented as

\eq{mod_action}{S = \int \left[ \frac{c^4}{16\pi G}f(R) + \mathcal{L}_M\right]. }

\noindent Now varying this modified action w.r.t  $g_{\mu\nu}$ with appropriate boundary condition gives a modified version of the field equation, which is given by \cite{Felice_Tsujikawa_10, Nojiri_Odintsov_Oikonomou_17,Nojiri:2010wj}

\eq{mod_Field_eq}{
\begin{split}
F(R)&G_{\mu \nu}+\frac{1}{2}g_{\mu \nu}[RF(R)-f(R)]\\ &-(\nabla_\mu \nabla_\nu-g_{\mu \nu}\Box)F(R) =\frac{8\pi G}{c^4}T_{\mu \nu},
\end{split}
}

\noindent where $F(R) = \frac{d}{dR}f(R)$,
$\Box$ is the d’Alembertian operator given by $\Box = \nabla^\mu\nabla_\mu$ and $\nabla_\mu$ is the covariant derivative. For $f(R) = R$, this equation reduces to the well-known Einstein field equation in GR.
 
Now for the vacuum solution the energy-momentum tensor vanishes, i.e. $T_{\mu\nu} = 0$, and the equation reduces to
\eq{mod_Field_eq_vacuum}{
\begin{split}
F(R)G_{\mu\nu}&+\frac{1}{2}g_{\mu\nu}\left[RF(R)-f(R)\right]\\ &-\left(\nabla_\mu\nabla_\nu-g_{\mu\nu}\Box\right)F(R) =0.
\end{split}
}
The trace of this equation is given by
\eq{Trace_eq}{RF(R)-2f(R)+3 \Box F(R) = 0.}
Substituting $f(R)$ from equation \Eref{Trace_eq} into equation \Eref{mod_Field_eq_vacuum}, we have
\eq{Trace_eq_2}{FR_{\mu\nu}-\nabla_\mu\nabla_\nu F= \frac{1}{4}g_{\mu\nu}\left(RF-\Box F\right).}

%%%%%%%%%%%%%%%%%%%%%%%%%%%%%%%%%%%%%%%%%%%%%%%%%%%
%%%%%%%%%%%%%%%%%%%%%%%%%%%%%%%%%%%%%%%%%%%%%%%%%%%
%%%%%%%%%  Non-rotating vacuum solution  %%%%%%%%%%
%%%%%%%%%%%%%%%%%%%%%%%%%%%%%%%%%%%%%%%%%%%%%%%%%%%
%%%%%%%%%%%%%%%%%%%%%%%%%%%%%%%%%%%%%%%%%%%%%%%%%%%

\section{Solution for a nonrotating black hole}

Here we briefly recapitulate a solution for a non-rotating black hole in 
$f(R)$-gravity obtained earlier \cite{Kalita_Bani}.
The vacuum solution of a spherically symmetric and static system can be written in the form of $g_{\mu \nu }=\mathrm{diag}\left(s\left(r\right),-p\left(r\right),\ -r^2,\ -r^2{{\mathrm{sin}}^2 \theta \ }\right)$. Now we assume that $F\left(R\right)$ has a form such that, $F\left(r\right)=1+B/r$. Hence, as $r\to \infty ,\ F\left(r\right)\to 1$, which generates the usual theory of GR. Note that $B\leq 0$ to guarantee the attractive nature of gravity \cite{Kalita_Bani}. Now from equation \Eref{Trace_eq} we have \cite{Mut_Vij_06} 

\eq{1}{2\frac{X'}{X}+r\frac{F'}{F}\frac{X'}{X}-2r\frac{F''}{F}=\ 0}

\noindent and

\eq{2}{-4s+4X-4rs\frac{F'}{F}+2r^2s' \frac{F'}{F} +2rs\frac{X'}{X}-r^2 s' \frac{X'}{X}+2r^2\ s''=\ 0,} 

\noindent where $X\left(r\right)=p\left(r\right)s\left(r\right)$.

Solving equations \Eref{1} and \Eref{2}, and applying the boundary condition $X\left(r\right)\to 1$ as $r\to \infty $, $X(r)$ can be found as \cite{Kalita_Bani}

\eq{ad}{X\left(r\right)=\frac{16r^4}{{\left(B+2r\right)}^4}.}

\noindent Putting equation \Eref{ad} in equation \Eref{2} we obtain the series solution for $s(r)$ (for $B\neq 0$) as

\eq{3}{
\begin{split}
s\left(r\right)={} \frac{-16+2BC_1+32{\mathrm{log} 2 }+\left(BC_2+8\right)i\pi}{2B^2}r^2\\+1+\frac{B\left(-24+BC_2\right)}{24r}
	+\frac{B^2-\frac{1}{16}B^3C_2}{r^2}\\+\frac{-B^3+\frac{11}{160}B^4C_2}{r^3}+\frac{188B^4-13B^5C_2}{192r^4}+ \dots,
\end{split}
}

\noindent where $C_1$ and $C_2$ are constants of integrations which can be 
obtained by arguing that the metric needs to behave as Schwarzschild metric at a large distance, which requires the coefficient of $r^2$ to vanish and coefficient of $1/r$ to be $-2$, which gives
\eq{}{C_2=\frac{24\left(B-2\right)}{B^2},} 
\eq{}{C_1=\ -8\frac{B\left(-1+{\mathrm{log} 4\ }\right)+\left(-3+2B\right)i\pi }{B^2}.}

\noindent Thus, the temporal component of the metric turns out to be
\eq{4}{
\begin{split}
g_{tt}=s\left(r\right)=\ 1-\frac{2}{r}-\frac{\left(-6+B\right)B}{2r^2}\\+\frac{B^2\left(-66+13B\right)}{20r^3}
-\frac{B^3\left(-156+31B\right)}{48r^4}\\+\frac{3B^4\left(-57+11B\right)}{56r^5}+O\left[r^{-6}\right].
\end{split}
}

\noindent Thus the radial component of the metric can be found as $g_{rr}=\ -p\left(r\right)$, where $p\left(r\right)=\ X(r)/s(r)$, and thus the power series solution takes the form as

\eq{5}{
\begin{split}
p\left(r\right)= 1+\frac{2-2B}{r}+\frac{(-1+B)(-4+3B)}{r^2}\\-\frac{\left(-2+B\right)\left(80+B\left(-160+83B\right)\right)}{20r^3}
\\+\frac{16-52B+\frac{1}{60}B^2\left(3732+B\left(-1917+338B\right)\right)}{r^4}
\\+\frac{32-128B+\frac{1008B^2}{5}-155B^3+\frac{6002B^4}{105}-\frac{6431B^5}{840}}{r^5}\\+O\left[r^{-6}\right].
\end{split}
}

%%%%%%%%%%%%%%%%%%%%%%%%%%%%%%%%%%%%%%%%%%%%%%%%%%%
%%%%%%%%%%%%%%%%%%%%%%%%%%%%%%%%%%%%%%%%%%%%%%%%%%%
%%%%%%%%%%%  Rotating vacuum solution  %%%%%%%%%%%%
%%%%%%%%%%%%%%%%%%%%%%%%%%%%%%%%%%%%%%%%%%%%%%%%%%%
%%%%%%%%%%%%%%%%%%%%%%%%%%%%%%%%%%%%%%%%%%%%%%%%%%%

\section{Rotating black hole}

%%%%%%%%%%%%%%%%%%%%%%%%%%%%%%%%%%%%%%%%%%%%%%%%%%%
%%%%  Newman-Janis Algorithm (NJA) revisited  %%%%%
%%%%%%%%%%%%%%%%%%%%%%%%%%%%%%%%%%%%%%%%%%%%%%%%%%%

\subsection{Revisiting basics of Newman-Janis Algorithm}

After the original discovery of the Kerr metric, Newman and Janis showed that the solution could be derived from the Schwarzschild solution by making an elementary transformation involved with complex numbers, assuming the black hole to be spinning. The spin (angular momentum per unit mass) of black hole comes into the solution as an arbitrary parameter. The static spherically symmetric metric and the line element could be written in the general form in $(+---)$ convention as \cite{Weinberg_1972} 

\eq{seed_line_element}{ds^2=s\left(r\right)dt^2-\mathrm{\ }p\left(r\right)dr^2-r^2\left(d{\theta }^2+{{\mathrm{sin}}^{\mathrm{2}} \theta \ }d{\phi }^2\right).}

\noindent In the null coordinates, this line element can be written, by 
advancing the time coordinate as $dt=du+\hat{f}dr$ and setting 
$\hat{f}={\left[s(r)/p(r)\right]}^{-\frac{1}{2}}$, as

%\eq{}{ds^2=s(r)\ du^2+2\ {\left[s\left(r\right)\ p\left(r\right)\right]}^{\frac{1}{2}}\ du\ dr -r^2\left(d{\theta }^2+{{\mathrm{sin}}^{\mathrm{2}} \theta \ }d{\phi }^2\right).}
\begin{eqnarray}
\nonumber
ds^2=s(r)\ du^2+2\ {\left[s\left(r\right)\ p\left(r\right)\right]}^{\frac{1}{2}}\ du\ dr\\ 
-r^2\left(d{\theta }^2+{{\mathrm{sin}}^{\mathrm{2}} \theta \ }d{\phi }^2\right).
\end{eqnarray}

\noindent Thus, the contravariant form of the metric can be written as

\eq{seed_metric}{g^{\mu \nu }=\left( \begin{array}{cccc}
0 & \ {\left[s\left(r\right)\ p\left(r\right)\right]}^{-\frac{1}{2}} & 0 & 0 \\ 
. & -\frac{1}{p\left(r\right)} & 0 & 0 \\ 
. & . & -\frac{1}{r^2} & 0 \\ 
. & . & . & -\frac{1}{r^2{{\mathrm{sin}}^{\mathrm{2}} \theta \ }} \end{array}
\right).}
Here ``$.$''s in equation \Eref{seed_metric} indicate that the metric is symmetric and will have the same elements as in the upper triangle. The contravariant form of the metric can be written so that it can be expressed in terms of its null tetrads \cite{NJA_Orig, NJA_1, NJA_Revisited} as

\eq{null_tetrads_general}{g^{\mu \nu }=l^{\mu }n^{\nu }+l^{\nu }n^{\mu }-m^{\mu }{\overline{m}}^{\nu }-m^{\nu }{\overline{m}}^{\mu },} 

\noindent where the null tetrads satisfy the conditions

\eq{null_tetrads_general_cond}{
\begin{split}
   &l_{\mu }l^{\mu }=m_{\mu }m^{\mu }=\ n_{\mu }n^{\mu }=0,
   \\&l_{\mu }n^{\mu }=\ -m_{\mu }{\overline{m}}^{\mu }=1,
   \\&l_{\mu }m^{\mu }=n_{\mu }m^{\mu }=0,
\end{split}
}
with the bar indicating the complex conjugate. 

Putting the elements of the metric from equation \Eref{seed_metric} 
to equation \Eref{null_tetrads_general}, along with equation \Eref{null_tetrads_general_cond}, the null tetrads are found to be

\eq{}{l^{\mu }={\delta }^{\mu }_1,}
\eq{22}{ n^{\mu }=\ -\frac{1}{2}\frac{1}{p\left(r\right)}{\delta }^{\mu }_1+\ {\left[s\left(r\right)\ p\left(r\right)\right]}^{-\frac{1}{2}}{\delta }^{\mu }_0,}
\eq{}{ m^{\mu }=\frac{1}{\sqrt{2}\ r}\left({\delta }^{\mu }_2+\frac{i}{{\mathrm{sin} \theta \ }}{\delta }^{\mu }_3\right).}
Then following NJA, we proceed by making a complex transformation as

\seq{}{
&u\to u'=u-ia \cos{\theta} \ ,
\\ &r\to r'=r+ia\cos{\theta},
\\ &\theta \to {\theta }'=\theta , 
\\&\phi \to {\phi }'=\phi.
}

\noindent By considering this as a complex rotation of the $\theta-\phi$ plane, the tetrads can be obtained as

\eq{}{l^{\mu }={\delta }^{\mu }_1,}
\eq{ad2}{n^{\mu }=\ -\frac{1}{2}\frac{1}{p\left(r,\theta \right)}{\delta }^{\mu }_1+{\left[s\left(r,\theta \right)\ p\left(r,\theta \right)\right]}^{-\frac{1}{2}}{\delta }^{\mu }_0,}

\begin{eqnarray}
	\nonumber
	m^{\mu }=\frac{1}{\sqrt{2}\ \left(r+ia{\mathrm{cos} \theta \ }\right)\ }&&\left({ia{\mathrm{sin} \theta \ }(\delta }^{\mu }_0-{\delta }^{\mu }_1)+{\delta }^{\mu }_2\right.\\ &&\left.+\frac{i}{{\mathrm{sin} \theta \ }}{\delta }^{\mu }_3\right).
\end{eqnarray}
%\seq{}{m^{\mu }=\frac{1}{\sqrt{2}\ \left(r+ia{\mathrm{cos} \theta \ }\right)\ }\left({ia{\mathrm{sin} \theta \ }(\delta }^{\mu }_0-{\delta }^{\mu }_1)+{\delta }^{\mu }_2+\frac{i}{{\mathrm{sin} \theta \ }}{\delta }^{\mu }_3\right).}
Note that $s(r,\theta )$ and $p(r,\theta )$ in equation \Eref{ad2} are completely different from $s\left(r\right)$ and $p\left(r\right)$ in equation \Eref{22} (and in equations \Eref{3} and \Eref{5}; also see \cite{azeg,azeg2,azeg3}). In fact, the new functions are functions of both $r$ and $\theta$, while the old ones are functions of only $r$.

From equation \Eref{null_tetrads_general}, the contravariant form of the metric is 
obtained as
\begin{widetext}
\eq{}{g^{\mu \nu }=\ \left( \begin{array}{cccc}
-\frac{a^2{{\mathrm{sin}}^{\mathrm{2}} \theta \ }}{\mathrm{\Sigma }} & {\left[s\left(r,\theta \right)\ p\left(r,\theta \right)\right]}^{-\frac{1}{2}}+\frac{a^2{{\mathrm{sin}}^{\mathrm{2}} \theta \ }}{\mathrm{\Sigma }} & 0 & -\frac{a}{\mathrm{\Sigma }} \\ 
. & -\frac{1}{p\left(r,\theta \right)}-\frac{a^2{{\mathrm{sin}}^{\mathrm{2}} \theta \ }}{\mathrm{\Sigma }} & 0 & \frac{a}{\mathrm{\Sigma }} \\ 
. & . & -\frac{1}{\mathrm{\Sigma }} & 0 \\ 
. & . & . & -\frac{1}{\mathrm{\Sigma }{{\mathrm{sin}}^{\mathrm{2}} \theta \ }} \end{array}
\right),}

\noindent where $\mathrm{\Sigma }=r^2+a^2{\mathrm{cos}}^{\mathrm{2}} \theta$. The inverse of this metric, i.e. its covariant form, is

\eq{}{g_{\mu \nu }=\ \left( \begin{array}{cccc}
s\left(r,\theta \right) & {\left[s\left(r,\theta \right)\ p\left(r,\theta \right)\right]}^{\frac{1}{2}} & 0 & a{{\mathrm{sin}}^{\mathrm{2}} \theta \ }\left({\left[s\left(r,\theta \right)\ p\left(r,\theta \right)\right]}^{\frac{1}{2}}-s\left(r,\theta \right)\right) \\ 
. & 0 & 0 & -a\ {\left[s\left(r,\theta \right)\ p\left(r,\theta \right)\right]}^{\frac{1}{2}}{{\mathrm{sin}}^{\mathrm{2}} \theta \ } \\ 
. & . & -\mathrm{\Sigma } & 0 \\ 
. & . & . & -{{\mathrm{sin}}^{\mathrm{2}} \theta \ }\left(\mathrm{\Sigma }+a^2{{\mathrm{sin}}^{\mathrm{2}} \theta \ }\left(2{\left[s\left(r,\theta \right)\ p\left(r,\theta \right)\right]}^{\frac{1}{2}}-s\left(r\right)\right)\right) \end{array}
\right).}
\end{widetext}

Now we redefine the coordinates $u$  and $\phi$ such that, $du=dt+g\left(r\right)\ dr\ $and $d\mathrm{\phi }=d\varphi +h(r)\ dr$, with $g$ and $h$ as

\eq{}{g\left(r\right)=\ -\frac{(p{\left(r,\theta \right))}^{\frac{1}{2}}\left(\mathrm{\Sigma }+a^2{{\mathrm{sin}}^{\mathrm{2}} \theta {\left[s\left(r,\theta \right)\ p\left(r,\theta \right)\right]}^{\frac{1}{2}}\ }\right)}{{\left(s\left(r\right)\right)}^{\frac{1}{2}}\left(\mathrm{\Sigma }+a^2{{\mathrm{sin}}^{\mathrm{2}} \theta \ }e^{2\mathrm{\lambdaup }\left(r,\theta \right)}\right)},}

\eq{}{h\left(r\right)=\ -\frac{a\ p\left(r\right)}{\mathrm{\Sigma }+a^2{{\mathrm{sin}}^{\mathrm{2}} \theta \ }p\left(r\right)},}

\noindent in a new coordinate system. This leads to all the non-diagonal 
elements, except $g_{\phi t}$, go to zero. 
This transforms the metric to Boyer-Lindquist coordinate system. Now putting $X\left(r,\theta \right)=p\left(r,\theta \right)s\left(r,\theta \right)$, the metric in this coordinate system takes the form

\begin{widetext}
\eq{6}{g_{\mu \nu }=\left( \begin{array}{cccc}
s\left(r,\theta \right) & 0 & 0 & a{{\mathrm{sin}}^{\mathrm{2}} \theta \ }\left({\left(X\left(r,\theta \right)\right)}^{\frac{1}{2}}-s\left(r,\theta \right)\right) \\ 
. & -\frac{\mathrm{\Sigma }}{\frac{\mathrm{\Sigma }\mathrm{\ }s\left(r,\theta \right)}{X\left(r,\theta \right)}\ +a^2{{\mathrm{sin}}^{\mathrm{2}} \theta \ }} & 0 & 0 \\ 
. & . & \mathrm{-}\mathrm{\Sigma } & 0 \\ 
. & . & . & -{{\mathrm{sin}}^{\mathrm{2}} \theta \ }\left(\mathrm{\Sigma }+a^2{{\mathrm{sin}}^{\mathrm{2}} \theta \ }\left(2{\left(X\left(r,\theta \right)\right)}^{\frac{1}{2}}-s\left(r\right)\right)\right) \end{array}
\right), }
\end{widetext}
which essentially leads to the counter part of rotating black hole of the metric in equation \Eref{seed_line_element}.

%%%%%%%%%%%%%%%%%%%%%%%%%%%%%%%%%%%%%%%%%%%%%%%%%%%
%% Transformation of specific functions under NJA %%
%%%%%%%%%%%%%%%%%%%%%%%%%%%%%%%%%%%%%%%%%%%%%%%%%%%

\subsection{Transformation of specific functions under NJA and modified Kerr metric}

\noindent Equipped with the knowledge of NJA, the angular momentum parameter can be easily incorporated in the non-rotating vacuum solution. For this we first proceed by noting that while we make the complex transformation, the coordinates $r$ and $u$ are complexified and a new parameter $a$ is introduced. 
However, since in the end one needs a real spacetime, a function $Q$ must remain real and so its changes are given as \cite{NJA_Revisited,Source_and_singularity}
\eq{}{Q\left(r\right)\to Q\left(r,\overline{r}\right), }
so that the functions $1/r^{2n}$ and $1/r^{2n+1}$ must be written as
\eq{}{\frac{1}{r^{2n}}\rightarrow \frac{1}{{\left(r\overline{r}\right)}^n},} 
\eq{}{\frac{1}{r^{2n+1}}= \frac{1}{r^{2n}}\frac{1}{r}\to \frac{1}{{\left(r\overline{r}\right)}^n}\frac{1}{2}\left(\frac{1}{r}+\frac{1}{\overline{r}}\right).}
Now suppose the function $Q\left(r,\overline{r}\right)$ has some terms of $\frac{1}{{\left(r\overline{r}\right)}^n}$ and $\frac{1}{{\left(r\overline{r}\right)}^{n}}\frac{1}{2}\left(\frac{1}{r}+\frac{1}{\overline{r}}\right)$ with at least one of them having a non-zero coefficient, then after the complex transformation of $u\to u'=u-ia \cos{\theta},\ \ r\to r'=r+ia\cos{\theta},\ \ \theta \to {\theta }'=\theta ,\ \ \phi \to {\phi }'=\phi$, the components of $Q\left(r,\overline{r}\right)$ will transform as

\eq{7}{\frac{1}{r^{2n}}\equiv \frac{1}{{\left(r\overline{r}\right)}^n}\to \frac{1}{\left[(r+ia\cos{\theta})(r-ia\cos{\theta})\right]^n}=\frac{1}{{\mathrm{\Sigma }}^n}}

\noindent and similarly

\seq{8}{&\frac{1}{r^{2n+1}}\equiv \frac{1}{{\left(r\overline{r}\right)}^n}\frac{1}{2}\left(\frac{1}{r}+\frac{1}{\overline{r}}\right)
\\&\to \frac{1}{{\left[(r+ia\cos{\theta})(r-ia\cos{\theta})\right]}^n}\frac{1}{2}\bigg(\frac{1}{r+ia\cos{\theta}}\\& \hspace{2cm}+\frac{1}{r-ia\cos{\theta}}\bigg)=\frac{r}{{\mathrm{\Sigma }}^{n+1}}.}

%\noindent where $\mathrm{\Sigma }=r^2+a^2{{\mathrm{cos}}^2 \theta .}$

 Thus, after the complex transformation, the function $Q(r)$ transforms to $Q(r,\theta )$.\footnote{$Q\left(r\right)$ and $Q(r,\theta)$ are not necessarily equal.}
Applying equations \Eref{7} and \Eref{8} to the functions $X\left(r\right)$, $s(r)$ and $p\left(r\right)$, we have 

\eq{9}{X\left(r,\theta \right)=\ \frac{{(r^2+a^2{{\mathrm{cos}}^{\mathrm{2}} \theta \ })}^2}{{\left({\left(\frac{B}{2}+r\right)}^2+a^2{{\mathrm{cos}}^{\mathrm{2}} \theta \ }\right)}^2},}

\meq{10}{s\left(r,\theta \right)=\ 1-\frac{2r}{r^2+a^2{{\mathrm{cos}}^{\mathrm{2}} \theta \ }}-\frac{\left(-6+B\right)B}{2\left(r^2+a^2{{\mathrm{cos}}^{\mathrm{2}} \theta \ }\right)}\cr+\frac{B^2\left(-66+13B\right)r}{20{\left(r^2+a^2{{\mathrm{cos}}^{\mathrm{2}} \theta \ }\right)}^2}
-\frac{B^3\left(-156+31B\right)}{48{\left(r^2+a^2{{\mathrm{cos}}^{\mathrm{2}} \theta \ }\right)}^2}\cr+\frac{3B^4\left(-57+11B\right)r}{56{\left(r^2+a^2{{\mathrm{cos}}^{\mathrm{2}} \theta \ }\right)}^3}+\dots,}

\meq{11}{p\left(r,\theta \right)=\ 1+\frac{(2-2B)r}{r^2+a^2{{\mathrm{cos}}^{\mathrm{2}} \theta \ }}+\frac{(-1+B)(-4+3B)}{r^2+a^2{{\mathrm{cos}}^{\mathrm{2}} \theta \ }}\cr-\frac{\left(-2+B\right)\left(80+B\left(-160+83B\right)\right)r}{20{\left(r^2+a^2{{\mathrm{cos}}^{\mathrm{2}} \theta \ }\right)}^2}
\cr+\frac{16-52B+\frac{1}{60}B^2\left(3732+B\left(-1917+338B\right)\right)}{{\left(r^2+{{\mathrm{cos}}^{\mathrm{2}} \theta \ }\right)}^2}
\cr+\frac{\left(32-128B+\frac{1008B^2}{5}-155B^3+\frac{6002B^4}{105}-\frac{6431B^5}{840}\right)r}{{\left(r^2+{{\mathrm{cos}}^{\mathrm{2}} \theta \ }\right)}^3}+\dots \ .}
Thus equations \Eref{6}, \Eref{9}, \Eref{10} and \Eref{11} essentially complete our development of the metric which is the asymptotically flat vacuum solution for a rotating black hole in a modified gravity. It can be easily seen that by setting $B=0$, we obtain the usual Kerr-metric. 
%This can be thought to be a verification of the above result, since setting $B=0$, the ``seed'' metric essentially becomes the Schwarzschild metric. Thus, after applying the NJ algorithm, it is expected to get the Kerr solution as a result.

%%%%%%%%%%%%%%%%%%%%%%%%%%%%%%%%%%%%%%%%%%%%%%%%%%%
%%%%%%%%%%%%%%%%%%%%%%%%%%%%%%%%%%%%%%%%%%%%%%%%%%%
%%%%%%%%%%%%  Source and Singularity  %%%%%%%%%%%%%
%%%%%%%%%%%%%%%%%%%%%%%%%%%%%%%%%%%%%%%%%%%%%%%%%%%
%%%%%%%%%%%%%%%%%%%%%%%%%%%%%%%%%%%%%%%%%%%%%%%%%%%

\section{ Source and Singularity}

\noindent From equation \Eref{6} we see that the metric becomes singular, when $s(r)$ or $p(r)$ becomes singular and that happens when $\mathrm{\Sigma }=0$, since $\mathrm{\Sigma }$ is present at the denominator in both. This shows that the metric becomes singular for \cite{Source_and_singularity}
\[r=0,\ \ \theta =\frac{\pi }{2}.\] 
This can be seen to be a geometric singularity by computing the curvature contraction $R_{\mu \nu \rho \lambda }R^{\mu \nu \rho \lambda }$. Further, it is an extended singularity, rather than `point -- like' singularity (as in Schwarzschild metric). 

Now defining local rectangular coordinate system
\begin{align*}
    x =& r \sin{\theta} \cos{\phi} + \alpha \sin{\theta}\sin{\phi},\\
    y =& r \sin{\theta}\sin{\phi} -\alpha \sin{\theta}\cos{\phi},\\
    z =& r\cos{\theta,}
\end{align*}
we immediately see that \(r=0,\ \ \theta =\pi/2\) corresponds to \(x^2 +y^2\ =\ \alpha^2\) and \( z=0\). Consequently, the physical singularity of the Kerr metric is a ring singularity. With the small $B$ approximation as made in section \ref{Analytical Approximation} below, the term involved with 
spin angular momentum transforms as $\alpha\approx a-1.5B$ (as will be clearer
in section \ref{Analytical Approximation} below), thus the radius and angular position as, respectively,

\eq{12}{\alpha = a - 1.5B,\ \ \theta =\frac{\pi }{2}.}

\noindent Therefore, the singularity can be seen to be on a circle of radius $\alpha$ around the origin in the $z=0$ plane. The solution can be considered to lie uniformly distributed on this circle, bounding an interior disc $\sqrt{x^2+y^2}\le\alpha$. This singularity signifies the presence of a rotating black hole and is termed as ring singularity.

%%%%%%%%%%%%%%%%%%%%%%%%%%%%%%%%%%%%%%%%%%%%%%%%%%%
%%%%%%%%%%%%%%%%%%%%%%%%%%%%%%%%%%%%%%%%%%%%%%%%%%%
%%%%%%%%%%%%%%%%%%%  Horizons  %%%%%%%%%%%%%%%%%%%%
%%%%%%%%%%%%%%%%%%%%%%%%%%%%%%%%%%%%%%%%%%%%%%%%%%%
%%%%%%%%%%%%%%%%%%%%%%%%%%%%%%%%%%%%%%%%%%%%%%%%%%%

\section{Horizons}

\noindent In addition to the ring-like curvature singularity, there are also additional coordinate singularities. Such coordinate singularities can be removed by suitable choice of coordinates, but they often underlie important physical phenomenon and have geometric description. Considering the Boyer-Lindquist coordinates for the metric 
given by \Eref{6}, we define $\mathrm{\Delta }$ as

\eq{Horizon_cond}{\mathrm{\Delta }=\mathrm{\Sigma }\frac{s\left(r,\theta \right)}{X\left(r,\theta \right)}+a^2{{\mathrm{sin}}^2 \theta, \ }} 

\noindent then $g_{rr}=-\mathrm{\Sigma }/\mathrm{\Delta }$, which becomes singular when $\mathrm{\Delta }=0$. The solution of $r$ for $\mathrm{\Delta }=0$ gives two real values $r_\pm$ of which $r_{-} \leq r_{+}$. These radii are referred to as outer $(r_{+})$ and inner $(r_{-})$ horizons; the former is called the event horizon
and the later one Cauchy horizon, and the region $r<r_{+}$ is referred to as the ‘interior’ of the black hole. It can be shown that the event horizon marks the point of no return. Now since $r_{-}$ lies inside the event horizon and no actual observer can have access to the interior of the 
event horizon, we avoid any discussion about the inner horizon $r_{-}$. 
%In the next section, we have numerically solved the event horizon $r_{H} \coloneqq r_{+}$, and in the later section, we have taken an approximate analytic solution to get the picture of the solution.

%%%%%%%%%%%%%%%%%%%%%%%%%%%%%%%%%%%%%%%%%%%%%%%%%%%
%%%%%%%%%%%%%%  Numerical Solution  %%%%%%%%%%%%%%%
%%%%%%%%%%%%%%%%%%%%%%%%%%%%%%%%%%%%%%%%%%%%%%%%%%%

\subsection{ Numerical Solution}

\noindent From equations \Eref{6}, \Eref{9}, \Eref{10} and \Eref{11} we obtain the metric components as a series solution and substituting them in equation \Eref{Horizon_cond} effectively gives $\Delta$.
Now $\mathrm{\Delta }=0$ has been numerically solved in order to obtain event 
horizon $r_H$ which is $r_+$. We will obtain an analytic approximation of the result in the next section. Tables \ref{Table1} and \ref{Table2} show $r_H$ for different $a$ and $B$ in the equatorial plane.

%%%%%%%%%%%%%%  Numerical Solution Table  %%%%%%%%%%%%%%%

\begin{table*}

    \centering

\begin{tabular}{|p{0.15in}|p{0.3in}|p{0.3in}|p{0.3in}|p{0.3in}|p{0.3in}|p{0.3in}|p{0.3in}|p{0.3in}|p{0.3in}|p{0.3in}|p{0.3in}|p{0.3in}|} \hline

${\boldsymbol{r}}_{\boldsymbol{H}}$& \multicolumn{12}{|c|}{$\boldsymbol{a}$\textbf{}} \\ \hline

\multirow{12}{4em}{$\boldsymbol{B}$}&  & \textbf{0} & \textbf{0.1} & \textbf{0.2} & \textbf{0.3} & \textbf{0.4} & \textbf{0.5} & \textbf{0.6} & \textbf{0.7} & \textbf{0.8} & \textbf{0.9} & \textbf{1} \\ \cline{2-13}

 & \textbf{0} & 2.00 & 1.99 & 1.98 & 1.95 & 1.92 & 1.87 & 1.80 & 1.71 & 1.60 & 1.44 & 1.00 \\  \cline{2-13}

 & \textbf{-0.1} & 2.15 & 2.14 & 2.13 & 2.11 & 2.07 & 2.02 & 1.96 & 1.88 & 1.78 & 1.64 & 1.42 \\  \cline{2-13}

 & \textbf{-0.2} & 2.30 & 2.29 & 2.28 & 2.26 & 2.22 & 2.18 & 2.12 & 2.05 & 1.95 & 1.83 & 1.66 \\  \cline{2-13}

 & \textbf{-0.3} & 2.45 & 2.44 & 2.42 & 2.41 & 2.37 & 2.33 & 2.28 & 2.21 & 2.12 & 2.01 & 1.86 \\  \cline{2-13}

 & \textbf{-0.4} & 2.60 & 2.59 & 2.58 & 2.55 & 2.52 & 2.48 & 2.43 & 2.37 & 2.29 & 2.19 & 2.06 \\  \cline{2-13}

 & \textbf{-0.5} & 2.73 & 2.74 & 2.72 & 2.70 & 2.67 & 2.63 & 2.59 & 2.52 & 2.45 & 2.36 & 2.24 \\  \cline{2-13}

 & \textbf{-1} & 3.46 & 3.45 & 3.45 & 3.43 & 3.41 & 3.37 & 3.33 & 3.28 & 3.23 & 3.16 & 3.07 \\  \cline{2-13}

 & \textbf{-1.5} & 4.17 & 4.16 & 4.15 & 4.14 & 4.12 & 4.09 & 4.06 & 4.02 & 3.97 & 3.91 & 3.84 \\  \cline{2-13}

 & \textbf{-2} & 4.86 & 4.86 & 4.85 & 4.84 & 4.82 & 4.79 & 4.76 & 4.73 & 4.68 & 4.63 & 4.58 \\  \cline{2-13}

 & \textbf{-2.5} & 5.55 & 5.45 & 5.54 & 5.53 & 5.51 & 5.49 & 5.46 & 5.43 & 5.39 & 5.34 & 5.29 \\  \cline{2-13}

 & \textbf{-3} & 6.23 & 6.25 & 6.22 & 6.21 & 6.19 & 6.17 & 6.15 & 6.12 & 6.08 & 6.04 & 5.99 \\ \hline

 \end{tabular}

    \caption{Numerical values of $r_H$ for varying $B$ and $a$.}

    \label{Table1}

\end{table*}

\begin{table}

    \centering

 \begin{tabular}{|p{0.15in}|p{0.3in}|p{0.3in}|p{0.3in}|p{0.3in}|p{0.3in}|} \hline

${\boldsymbol{r}}_{\boldsymbol{H}}$\textbf{} & \multicolumn{5}{|c|}{$\boldsymbol{a}$\textbf{}} \\ \hline

\multirow{12}{4em}{$\boldsymbol{B}$}& & \textbf{1.1} & \textbf{1.2} & \textbf{1.3} & \textbf{1.4}  \\ \cline{2-6}

 & \textbf{0} & - & - & - & - \\  \cline{2-6}

 & \textbf{-0.1} & - & - & - & - \\  \cline{2-6}

 & \textbf{-0.2} & 1.24 & - & - & - \\  \cline{2-6}

 & \textbf{-0.3} & 1.63 & - & - & - \\  \cline{2-6}

 & \textbf{-0.4} & 1.87 & 1.44 & - & - \\   \cline{2-6}

 & \textbf{-0.5} & 2.08 & 1.84 & - & - \\  \cline{2-6}

 & \textbf{-0.6} & 2.27 & 2.08 & - & - \\  \cline{2-6}

 & \textbf{-0.7} & 2.46 & 2.29 & 2.03 & - \\  \cline{2-6}

 & \textbf{-0.8} & 2.63 & 2.48 & 2.28 & - \\  \cline{2-6}

 & \textbf{-0.9} & 2.80 & 2.67 & 2.49 & 2.20 \\  \cline{2-6}

 & \textbf{-1} & 2.97 & 2.85 & 2.69 & 2.46 \\

\hline

\end{tabular}

    \caption{Numerical values of $r_H$ for varying $B$ and $a$, with $a>1$.}

    \label{Table2}

\end{table}

%\Todo{This table does not look that good}
%%%% Image %%%% 
\begin{figure}[h]
%    \centering
    \includegraphics[width=0.48\textwidth]{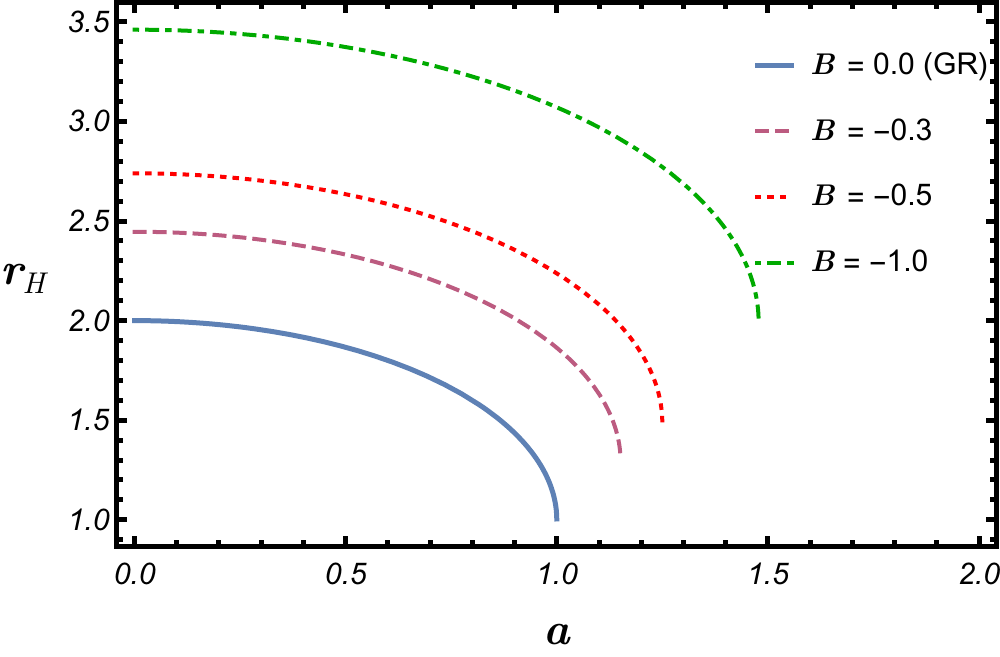}
    \caption{Variation of event horizon $r_H$ as a function of spin of black hole $a$ for different modified 
	gravity parameter $B$.}
    \label{Num_r_H_VS_a_var_B_FIG}
\end{figure}
%%%% Image %%%%

%%%% Image %%%% 
\begin{figure}[h]
%\centering
    \includegraphics[width=0.48\textwidth]{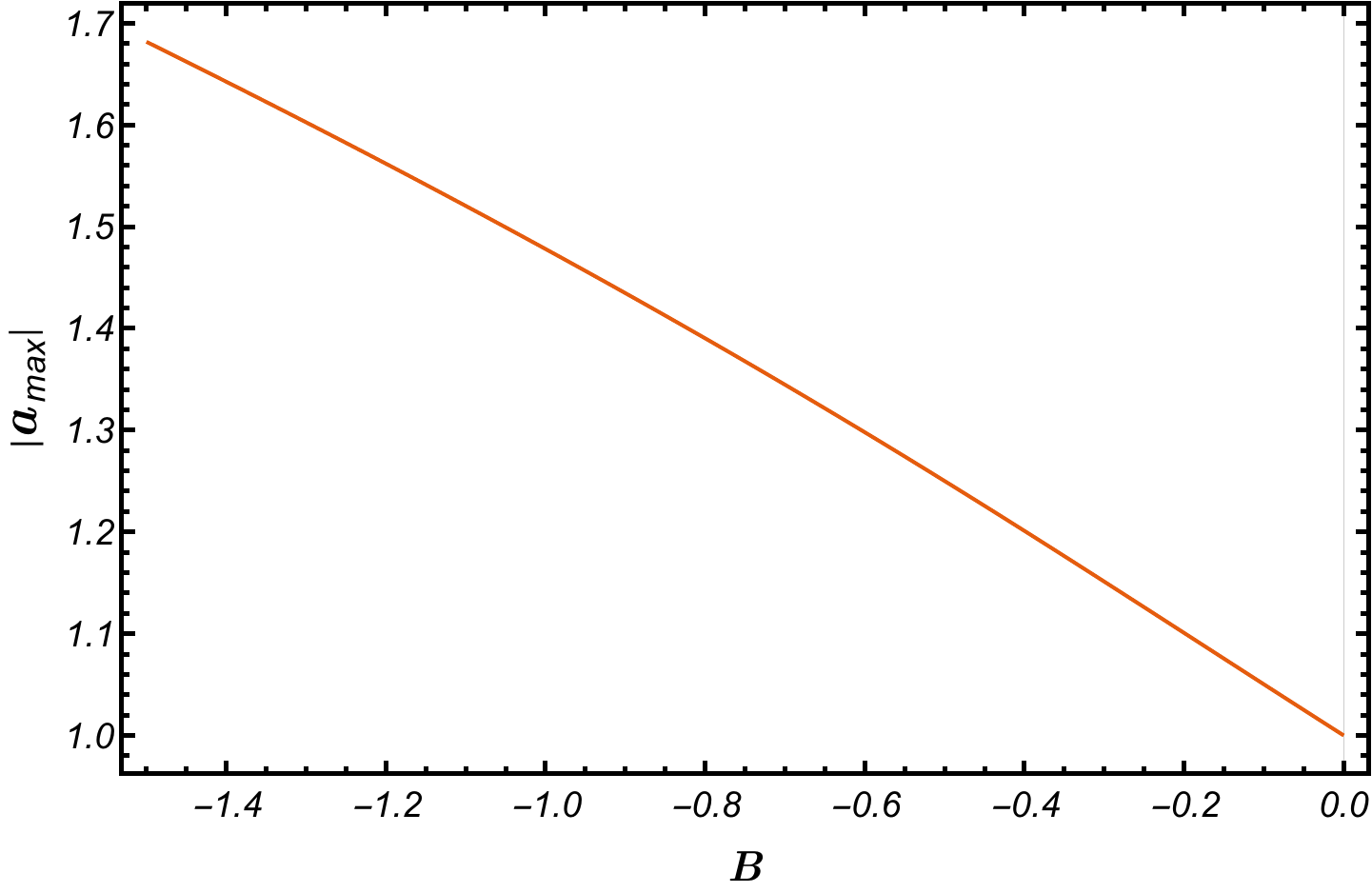}
    \caption{Variation of maximum possible spin of black hole $a_{max}$ as a function of modified
	gravity parameter $B$.}
    \label{a_max_VS_B_Fig}
\end{figure}
%%%% Image %%%%

Tables \ref{Table1} and \ref{Table2} show that $r_H$ monotonically increases with the increase of $|B|$ and monotonically decreases with the increase of $a$. From Table \ref{Table2} and Figure \ref{Num_r_H_VS_a_var_B_FIG} it can be seen that 
unlike in Kerr metric, $\left|a_{max}\right|>1$ is allowed due to $B<0$. The variation of maximum $a$, i.e.
$a_{max}$, for varying $B$ is shown in Figure \ref{a_max_VS_B_Fig}. It can be seen from the Figure \ref{a_max_VS_B_Fig} that $|a_{max}|$ varies almost linearly with $B$. Exploring and interpreting these results with the exact solutions is beyond the scope of this work. We will look at an analytic approximation of the above feature and report the result in the next section, where we will calculate $|a_{max}|$. We will confirm that 
indeed $|a_{max}|$ is allowed to be greater than unity in modified gravity and also varies approximately linearly with $B$.

%%%%%%%%%%%%%%%%%%%%%%%%%%%%%%%%%%%%%%%%%%%%%%%%%%%
%%%%%%%%%%%  Analytical Approximation  %%%%%%%%%%%%
%%%%%%%%%%%%%%%%%%%%%%%%%%%%%%%%%%%%%%%%%%%%%%%%%%%

\subsection{ Analytical Approximation} \label{Analytical Approximation}

\noindent In order to assure the possibility of analytical solutions, we consider very small modifications to GR and hence we take $B/r\ll 1$. Thus we take only terms up to $r^{-2}$, the functions $s(r,\theta )$ and $p(r,\theta )$ can then be written as

\begin{eqnarray}
	\nonumber
	s\left(r,\theta \right)=\ 1-\frac{2r}{r^2+a^2{{\mathrm{cos}}^{\mathrm{2}} \theta \ }}&&-\frac{\left(-6+B\right)B}{2\left(r^2+a^2{{\mathrm{cos}}^{\mathrm{2}} \theta \ }\right)}\\&&+O[r^{-3}],
\label{app_s_r_theta}
\end{eqnarray}
%\eq{app_s_r_theta}{s\left(r,\theta \right)=\ 1-\frac{2r}{r^2+a^2{{\mathrm{cos}}^{\mathrm{2}} \theta \ }}-\frac{\left(-6+B\right)B}{2\left(r^2+a^2{{\mathrm{cos}}^{\mathrm{2}} \theta \ }\right)}+O[r^{-3}],}

\meq{app_p_r_theta}{p\left(r,\theta \right)=\ 1+\frac{(2-2B)r}{r^2+a^2{{\mathrm{cos}}^{\mathrm{2}} \theta \ }}\\+\frac{(-1+B)(-4+3B)}{r^2+a^2{{\mathrm{cos}}^{\mathrm{2}} \theta \ }}+O\left[r^{-3}\right].}

\noindent Taking terms upto $r^{-2}$, in Boyer-Lindquist coordinate system, the metric can be recast from equations \Eref{6}, (\ref{app_s_r_theta}), \Eref{app_p_r_theta} and taking further $B\ll 1$ and having $X\approx1$, the nonzero component of the metric comes out to be

\eq{}{g_{\mu \nu }=\left( \begin{array}{cccc}
1-\frac{2r}{\mathrm{\Sigma }}-\frac{\beta}{\mathrm{\Sigma }}& 0 & 0 & \frac{2a{{\mathrm{sin}}^{\mathrm{2}} \theta}}{\Sigma}(2r+\beta) \\ 
. & -\frac{\mathrm{\Sigma }}{\Delta} & 0 & 0 \\ 
. & . & \mathrm{-}\mathrm{\Sigma } & 0 \\ 
. & . & . & -{{\mathrm{sin}}^{\mathrm{2}} \theta \ }\left( r^2 + a^2 +  \frac{a^2\sin{\theta}(2r+\beta)}{\Sigma}\right) \end{array}
\right),}

\noindent where $\Delta \approx r^2 + a^2 -2r -\beta$, and $\beta=\left(-6+B\right)B/2\approx-3B$. \footnote{Note, $B\le 0\to \beta\ge 0$ as $B=-\beta/3$.}
Thus the line element is of the form

\meq{}{ds^2 = \left(1-\frac{2r}{\mathrm{\Sigma }}-\frac{\beta}{\mathrm{\Sigma }}\right) dt^2 + \left(\frac{4a{{\mathrm{sin}}^{\mathrm{2}}  \theta}}{\Sigma}(2r+\beta)\right) dt d\varphi 
\\-\frac{\Sigma}{\Delta}dr^2 - \Sigma d\theta^2 -{{\mathrm{sin}}^{\mathrm{2}} \theta \ }\left( r^2 + a^2+ \frac{a^2\sin{\theta}(2r+\beta)}{\Sigma}\right)d\varphi^2 .}

\noindent This line element matches exactly with the results of black hole
theories with higher-dimensional branes \cite{Charged_BH_Brane, BH_Brane}. This shows that the work presented here gives a more general metric and includes the results from higher-dimensional branes. The effects of higher-dimensional branes come from a specialized case where the modification to gravity has been taken to be very small.

Now to find the horizons in this case, the equation $\mathrm{\Delta }=0$ has to be solved which approximately becomes, from equation \Eref{Horizon_cond},

\eq{}{\mathrm\Delta={\Sigma }\left(1-\frac{2r}{\mathrm{\Sigma }}-\frac{\beta}{\mathrm{\Sigma }}\right)+a^2{{\mathrm{sin}}^2 \theta \ }\approx0,}

\noindent which gives
\begin{equation}
\Delta=r^2-2r+\left(a^2-\beta\right)=0.
\label{Del}
\end{equation}

\noindent Thus, to the first order in $B$, we obtain $\mathrm{\Delta }=r^2+a^2-2r-\beta$. Now solving the quadratic equation (\ref{Del}) gives two three-surfaces of constant $r$ as
\[r_{\pm }=1\pm \sqrt{1-\left(a^2-\beta \right)}.\] 
These surfaces give the outer $(r_+)$ and inner $\left(r_-\right)$ horizons. Thus, the event horizon takes the form as
\eq{13}{r_H=r_+=1+\sqrt{1-a^2+\beta}.}
%%%% Image %%%% 
\begin{figure}[h]
%\centering
    \includegraphics[width=0.48\textwidth]{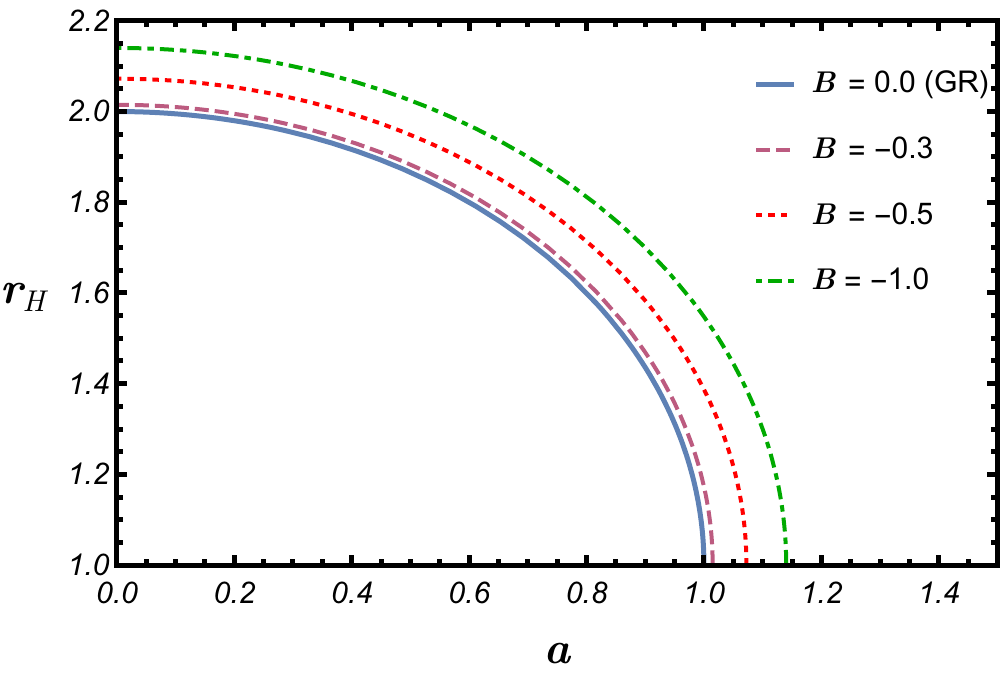}
    \caption{Approximate analytic solution for $r_H$ with the change of $a$ for different $B$.}
    \label{fig:1}
\end{figure}
%%%% Image %%%%

\noindent It can be easily seen that by setting  $B=0$, we recover the well-known results of the event horizon in Kerr metric, $r_{{H}_0}=1+\sqrt{1-a^2}\ $, which confirms the validity of analytical solutions.

Figure \ref{fig:1} shows how $r_H$ varies with $a$ based on analytical 
approximate solution.
It can be seen from Table \ref{Table1} that for $a=0$ the results  match quite well with the analytical results presented here. However, as $|B|$ increases, the value deviates a lot from the actual solution, which is because we have taken only terms up to $r^{-2}$ in $s\left(r\right)$ and $p\left(r\right)$ in analytical calculation. 
Quantitatively, when $B\approx -0.1$, very small compared to $r$, the numerical solution matches with the approximate analytical solution; thus, the analytical approximation is valid for the $B\ge -0.1$ realm, so that
\[{\left(r_{H_{analytical}}\right)}_{B\ge -0.1}\approx {\left(r_{H_{numerical}}\right)}_{B\ge -0.1}.\] 
From equation \Eref{13}, for $r_H$ to be real we must have
\[1-\left(a^2-\beta \right)\ge 0,\] 
\[\to \left|a\right|\le \sqrt{1+\beta }.\]
Thus,
\eq{a_max_eq}{|a_{max}| = \sqrt{1+\beta },}
\eq{approx_a_max_eq}{|a_{max}|\approx 1 + \beta/2 \approx 1-1.5B\geq|a_{max}|_{_{kerr}}.}
From equation \Eref{a_max_eq} the maximum value of $|a_{max}|$ obtained to be different from that obtained from Kerr metric and because $\beta \geq 0$, black holes can have spin parameter of value more than unity, i.e. $\left|a\right|\ge 1$. 
%Thus, we can say that objects can have a spin of value more than unity in modified gravity. 
The linear dependence of spin on modified gravity parameter can also be seen from equation \Eref{approx_a_max_eq} which nearly matches with Figure \ref{a_max_VS_B_Fig}.

Interestingly, this approximate analytical solution matches exactly with the 
Kerr-Newman metric if we replace $\beta$ with $-Q^2$, where $Q$ is the charge of the black hole. However, we know that the Kerr-Newman solution is a vacuum solution 
of the Einstein's field equation when the integrand of action is a scalar curvature 
(Ricci scalar) dependent on the parameters $M$, $a$ and $Q$. Hence, this
approximate solution due to the perturbative correction to GR can be treated
as the solution of Einstein's field equation itself with appropriate 
redefinition of the action and parameter(s). 
However, in general the solution ($g_{\mu\nu}$) obtained in \S IV can be understood
as the one corresponding to an appropriate choice of $f(R)$ and then $F(R)$ 
satisfying equation \Eref{mod_Field_eq_vacuum}.

%%%%%%%%%%%%%%%%%%%%%%%%%%%%%%%%%%%%%%%%%%%%%%%%%%%
%%%%%%%%%%%%%%%%%%%%%%%%%%%%%%%%%%%%%%%%%%%%%%%%%%%
%%%%%%%%%%  Orbits in equatorial plane  %%%%%%%%%%%
%%%%%%%%%%%%%%%%%%%%%%%%%%%%%%%%%%%%%%%%%%%%%%%%%%%
%%%%%%%%%%%%%%%%%%%%%%%%%%%%%%%%%%%%%%%%%%%%%%%%%%%

\section{ Orbits in equatorial plane}\label{Orbits in equatorial plane}

\noindent Due to the source having an angular momentum, the system's geometry is no longer spherical and is only axisymmetric. Only the components of the angular momentum along the symmetry axis are conserved. There are orbits confined to the equatorial plane ($\theta =\pi/2$), but the general orbit is not necessarily on the plane. However, to present a manageable solution, we consider the equatorial plane in this section. Thus, from 
equations \Eref{6}, \Eref{9}, \Eref{10} and \Eref{11} we can construct two Killing vectors corresponding to energy and angular momentum. The energy arises from the timelike Killing vector $K_{\mu }={\partial }_t$, and the Killing vector whose conserved quantity is the magnitude of the angular momentum is given by $L={\partial }_{\varphi }$. Thus, we can construct the conserved quantities as $E$ and $L$ as the conserved energy per unit mass and angular momentum per unit mass along the symmetry axes, which can be expressed as \cite{Hartle_Gravity_book}
\eq{killing_vector_1}{E=-K_{\mu }\ u^{\mu }}
and 
\eq{killing_vector_2}{L=L_{\mu }u^{\mu }.}
Now by inspecting the metric we have
\eq{e_energy}{E=-g_{tt}u^t-g_{t\varphi }u^{\varphi },}
\eq{l_ang_momentum}{L=g_{t\varphi }u^t+g_{\varphi \varphi }u^{\varphi }.}

\noindent These equations \Eref{killing_vector_1}, \Eref{killing_vector_2}, \Eref{e_energy} and \Eref{l_ang_momentum} can be solved for $u^t$ and $u^{\varphi }$ to find

\eq{u_t}{u^t=\frac{1}{\mathrm{\Delta }}\left(g_{\varphi \varphi }E+\ g_{t\varphi }L\right),}
\eq{u_phi}{u^{\varphi }=-\frac{1}{\mathrm{\Delta }}\left(g_{tt}L+g_{t\varphi }E\right),}
where $\mathrm{\Delta }={\left(g_{t\varphi }\right)}^2-g_{\varphi \varphi }g_{tt}$.

%%%%%%%%%%%%%%%%%%%%%%%%%%%%%%%%%%%%%%%%%%%%%%%%%%%
%%%%%%%%%%%  Marginally bound circular orbits  %%%%%%%%%%%%
%%%%%%%%%%%%%%%%%%%%%%%%%%%%%%%%%%%%%%%%%%%%%%%%%%%

\subsection{ Marginally bound circular orbit}\label{Marginally stable orbits}

\noindent From normalization condition of four-velocity $\boldsymbol{u}\boldsymbol{\cdot }\boldsymbol{u}=1$, together with $u^{\theta }=0$, we obtain a radial equation for $u^r=dr/d\tau$ as
\eq{u_normalize}{g_{tt}{\left(u^t\right)}^2+g_{rr}{\left(u^r\right)}^2+2g_{t\varphi }u^tu^{\varphi }+g_{\varphi \varphi }{\left(u^{\varphi }\right)}^2=1.} 
Thus equations \Eref{u_t}, \Eref{u_phi} and \Eref{u_normalize} essentially calculate $u^r$ as a function of $E$, $L$, $r$, $a$ and $B$. The effective potential can now be defined as \cite{Hartle_Gravity_book,STbook} 
\eq{}{V_{eff}\left(E,L,r,a,B\right):=r^3{\left(u^r\right)}^2.}
Now for circular orbits we must have the radial velocity to vanish and hence the effective potential must vanish. Thus for equilibrium condition, we must have an extremum in $V_{eff}$. Therefore, we obtain the relations
\eq{14}{V_{eff}=0,\ \ \frac{\partial V_{eff}}{\partial r}=0.}

\begin{figure}[H]
%\centering
    \includegraphics[width=0.48\textwidth]{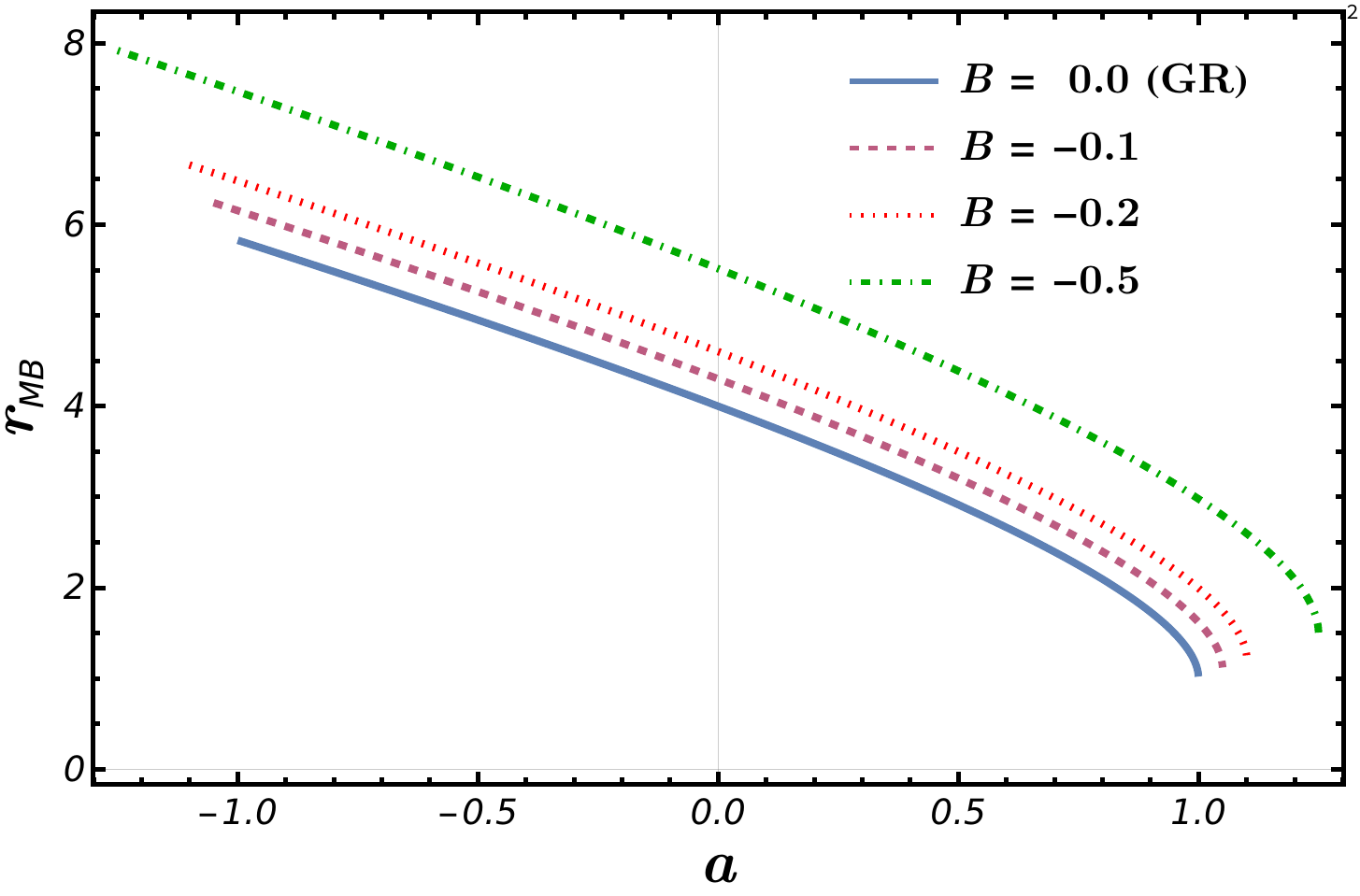}
    \caption{Variation of marginally bound orbit as a function of spin of black hole for different $B$.}
    \label{r_MB_r_Vs_a}
\end{figure}

\noindent It can be shown that unbound circular orbits have $E>1$. Given an infinitesimal outward perturbation, a particle in such an orbit will escape infinity. Bound orbits exist for $r>r_{mb}$, where $r_{mb}$ is the radius of the marginally bound circular orbit with  $E=1.$ Thus, solving
equation \Eref{14} with condition $E=1$, we obtain the value of $r=r_{mb}$. From Figure \ref{r_MB_r_Vs_a} the effect of $B$ on $r_{mb}$ can be seen, and that $r_{mb}$ increases with increasing $|B|$ for a fixed $a$, and $r_{mb}$ decreases with the increase of $a$ for a fixed $B$. It also can be seen that setting $B=0$ gives the same results as in GR.

%%%% Image %%%% 
%%%% Image %%%%

%%%%%%%%%%%%%%%%%%%%%%%%%%%%%%%%%%%%%%%%%%%%%%%%%%%
%%%%%%%  Innermost stable circular orbits  %%%%%%%%
%%%%%%%%%%%%%%%%%%%%%%%%%%%%%%%%%%%%%%%%%%%%%%%%%%%

\subsection{ Innermost stable circular orbit}\label{ISCO}

\noindent To find the innermost stable circular orbit, we opt for the same $V_{eff}$ as defined in section \ref{Marginally stable orbits}. Since we are considering circular orbits, equation \Eref{14} is still valid. All the bound circular orbits are not stable. For stability condition, we must have the condition
\eq{ext}{\frac{{\partial }^2V_{eff}}{\partial r^2}\le 0.} 

\begin{figure}[H]
%\centering
    \includegraphics[width=0.48\textwidth]{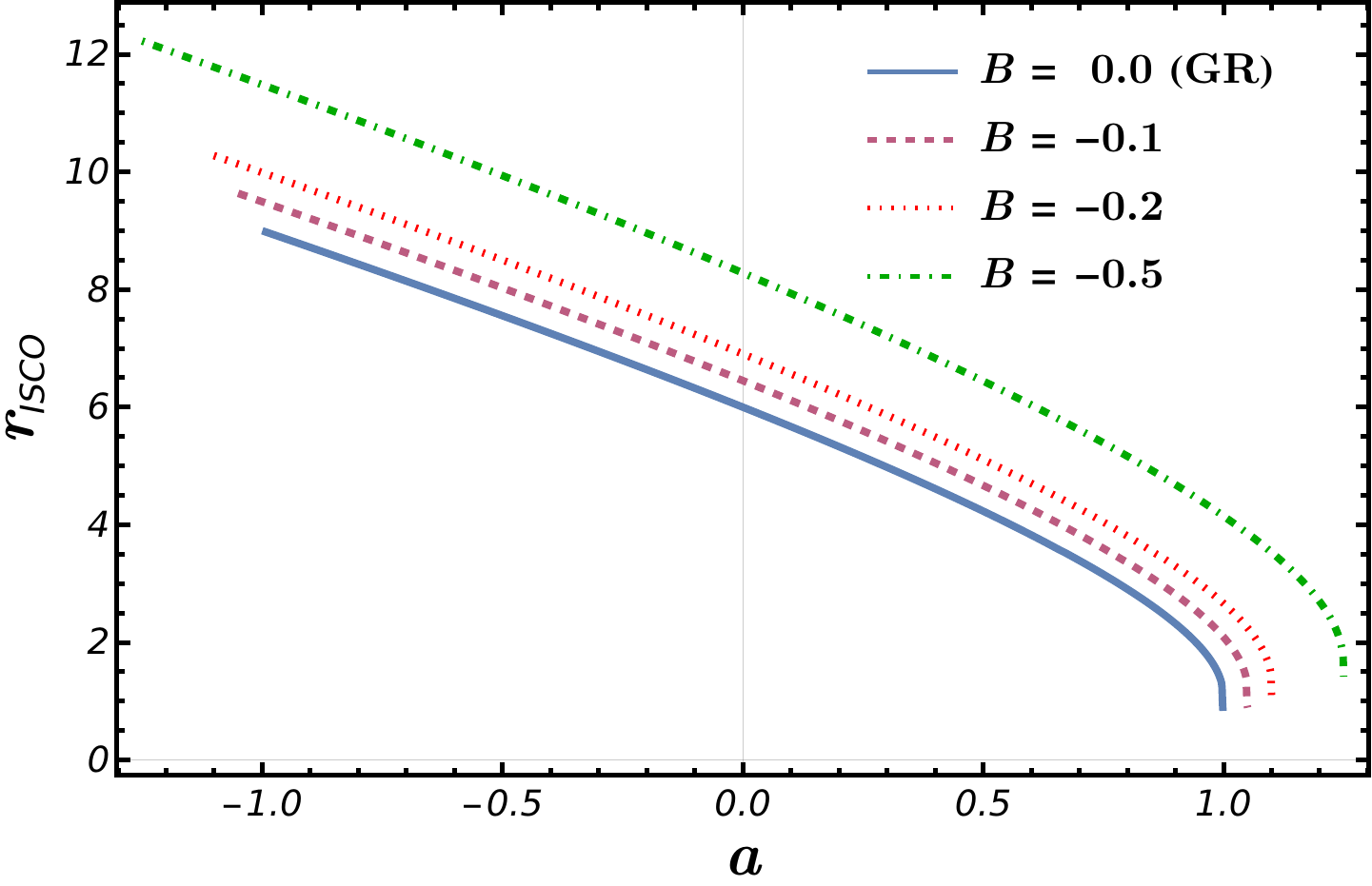}
    \caption{Variation of marginally stable circular orbit as a function of spin of black hole for 
	different $B$.}
    \label{r_ISCO_r_Vs_a}
\end{figure}

\noindent Now, the minimum radius (innermost orbit) that satisfies equations \Eref{14} and \Eref{ext} is termed as Innermost Stable Circular Orbit (ISCO) and the radius named as $r_{ISCO}$. 
Numerically solving these three equations simultaneously we obtain the variation of $r_{ISCO}$ shown
in Figure \ref{r_ISCO_r_Vs_a}. Similar to the case of  $r_{mb}$, here we see $r_{ISCO}$ increases with increasing $|B|$ for a fixed $a$, and $r_{ISCO}$ decreases with the increase of $a$ for a fixed $B$. Also, it can be easily verified that as $B=0$, the results of GR are preserved.
%%%% Image %%%% 
%%%% Image %%%%

%%%%%%%%%%%%%%%%%%%%%%%%%%%%%%%%%%%%%%%%%%%%%%%%%%%
%%%%%%%%%%%%%%%%%%%%%%%%%%%%%%%%%%%%%%%%%%%%%%%%%%%
%%%  Epicyclic frequency in modified gravity  %%%%%
%%%%%%%%%%%%%%%%%%%%%%%%%%%%%%%%%%%%%%%%%%%%%%%%%%%
%%%%%%%%%%%%%%%%%%%%%%%%%%%%%%%%%%%%%%%%%%%%%%%%%%%

\section{Epicyclic frequency in modified gravity}\label{Epicyclic frequency in modified gravity}
%%%% Image %%%% 
\begin{figure*}[t]
	\begin{subfigure}[b]{0.49\textwidth}
         \centering
         \includegraphics[width=0.8\textwidth]{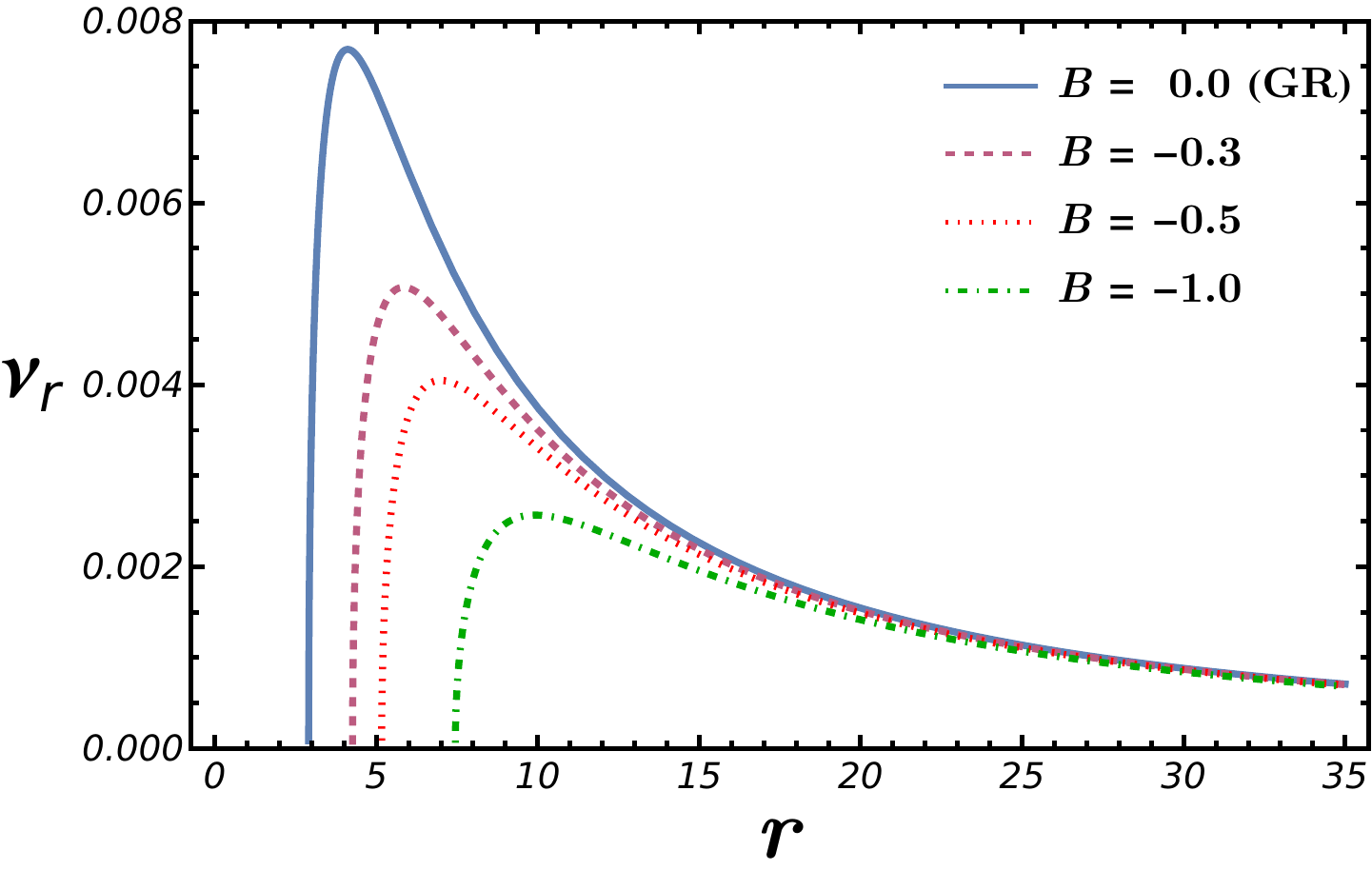}
	\caption{radial epicyclic frequencies $\nu_r$}
    		\label{nu_r_for_a_1}
     \end{subfigure}
    \hfill
     \begin{subfigure}[b]{0.49\textwidth}
         \centering
         \includegraphics[width=0.8\textwidth]{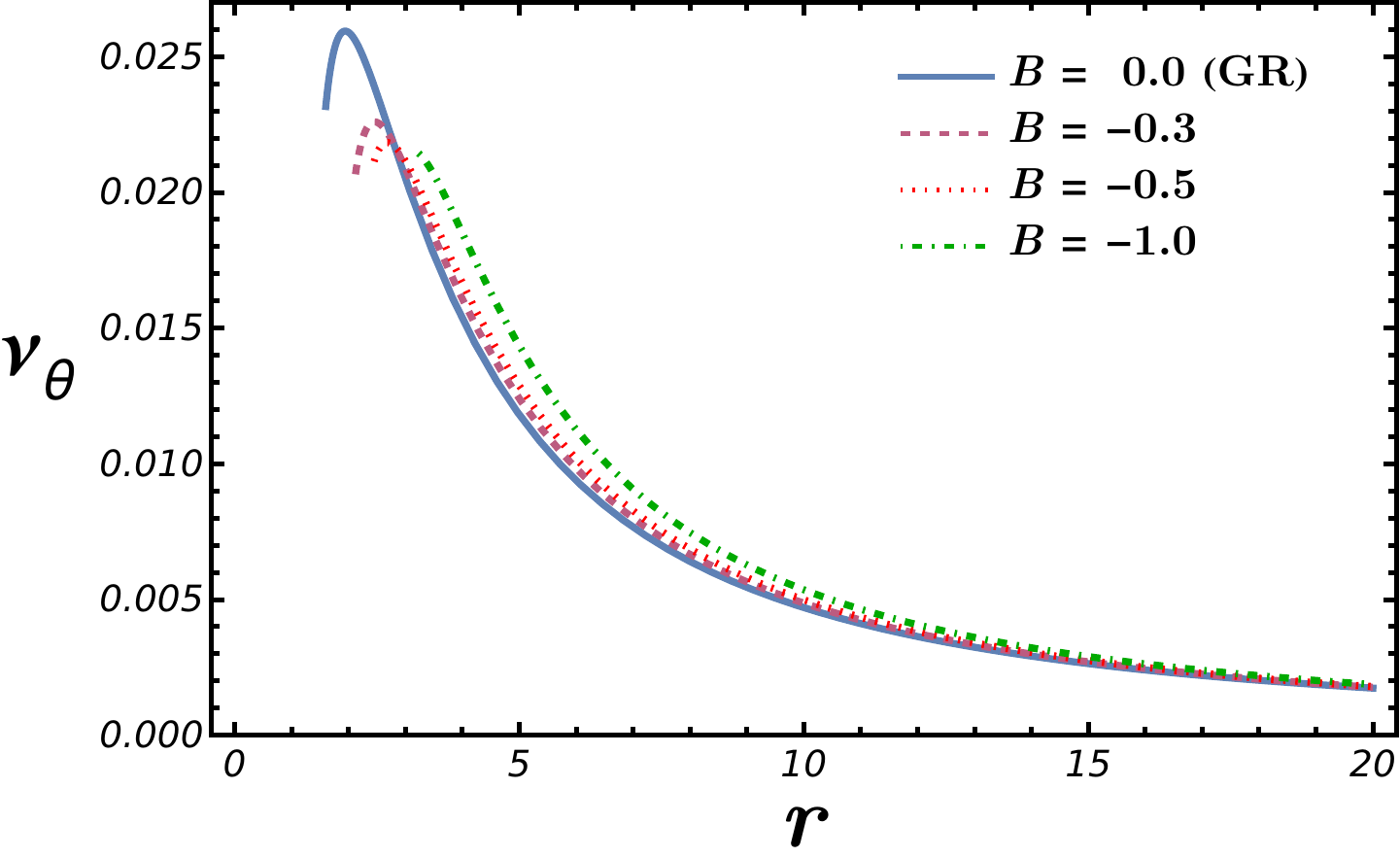}
	\caption{vertical epicyclic frequencies $\nu_\theta$}
    		\label{nu_theta_for_a_1}
     \end{subfigure}
     \newline
	\newline
     \begin{subfigure}[b]{0.49\textwidth}
         \centering
         \includegraphics[width=0.8\textwidth]{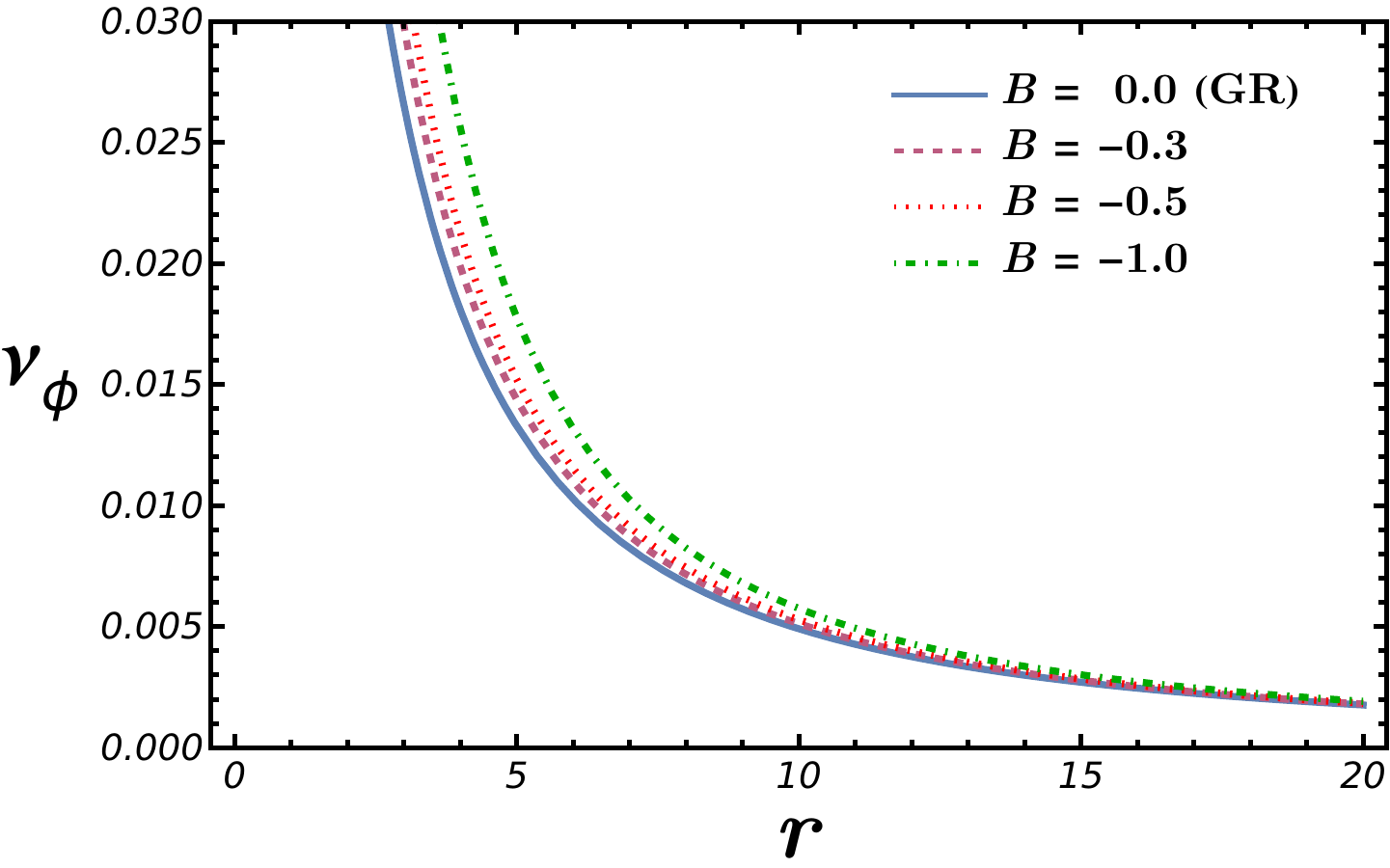}
	\caption{orbital angular frequency $\nu_\varphi$}
    		\label{nu_phi_for_a_1}
     \end{subfigure}
 \begin{subfigure}[b]{0.49\textwidth}
         \centering
         \includegraphics[width=0.8\textwidth]{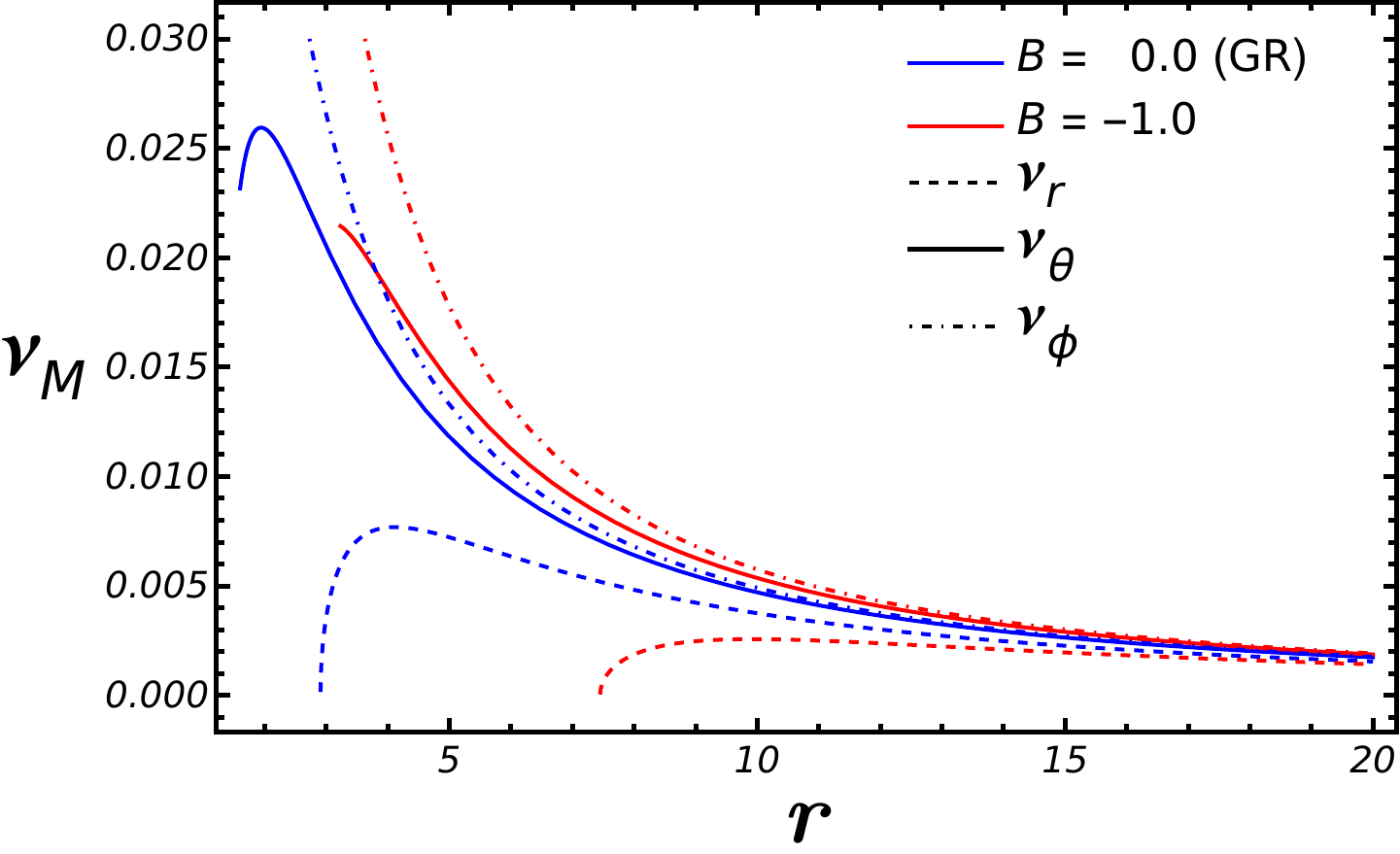}
	\caption{Comparison of various oscillation frequencies}
    		\label{nu_all_for_a_1}
     \end{subfigure}
        \caption{Profiles of the Keplerian frequency $\nu_\varphi$ and the epicyclic frequencies $\nu_r$ (radial) and $\nu_\theta$ (vertical) in the modified theory with spin parameter set to $a = 0.8$. The modified gravity parameter $B$ has been set to 0.0, -0.3, -0.5, -1.0 in (a), (b), and (c). In (d) $B=0$ and 
-1.0 for bottom and top curves respectively for $\nu_\varphi$ and $\nu_\theta$,
while for top and bottom curves for $\nu_r$.}
        \label{four graphs}
\end{figure*}
%%%% Image %%%%
In this section we will briefly describe the derivation of epicyclic oscillation frequencies for the stationary, axisymmetric metric from the effective potential for circular geodesics, depicting the spacetime around a rotating black hole. From equations \Eref{6}, \Eref{9}, \Eref{10} and \Eref{11} the line element can essentially be expressed as
\eq{line_element}{ds^2 = g_{tt}dt^2+2g_{t\varphi}dtd\varphi+g_{\varphi\varphi}d\varphi^2+g_{rr}dr^2+g_{\theta\theta}d\theta^2,} 
with $g_{\mu\nu}$ as a function of $r$ and $\theta$ and a symmetry along $\phi$
and $t$. It is most straightforward to obtain the epicyclic frequencies for a metric that can be expressed in this form. Epicyclic frequencies originate from the the relaxation of the circular orbits under external perturbation and it must be that this frequencies solely depend on the structure of the spacetime. 
%For example, the particles in the accretion disks are moving around the central object with varied radius, owing to two killing vectors mentioned in equations \Eref{killing_vector_1} and \Eref{killing_vector_2}, that arises because of certain symmetries in the line element. 

Now the similar normalization condition as in equation \Eref{u_normalize} along with equations \Eref{u_t} and \Eref{u_phi} but without a fixed $\theta$, hence with $u^\theta$, can be rewritten as 
\eq{u_normalize_new}{g_{rr}\left(u^r\right)^2+g_{\theta\theta}\left(u^\theta\right)^2 = \mathcal{V}_{eff},}
where the effective potential can be defined as
\eq{effective_potential_new}{\mathcal{V}_{eff} = \frac{\left(E^2-g_{tt}\right)g_{\varphi\varphi}+\left(2LE+g_{t\varphi}\right)g_{t\varphi}+L^2g_{tt}}{\left(g_{t\varphi}^2-g_{tt}g_{\varphi\varphi}\right)\Delta}.}
For circular orbits in the equatorial plane we have $u^r = u^\theta = 0$, which implies $\mathcal{V}_{eff} = 0$, and $\dot{u}^r = \dot{u}^\theta = 0$ give $\partial_r\mathcal{V}_{eff} = \partial_\theta\mathcal{V}_{eff} = 0$. From these 
three conditions $E$ and $L$ can be obtained as \cite{c_Bambi_12}
\eq{}{E = -\frac{g_{tt}+\Omega g_{t\varphi}}{\sqrt{-g_{tt}-2g_{t\varphi}\Omega-g_{\varphi\varphi}\Omega^2}},}

\eq{}{L = -\frac{g_{t\varphi}+\Omega g_{\varphi\varphi}}{\sqrt{-g_{tt}-2g_{t\varphi}\Omega-g_{\varphi\varphi}\Omega^2}}}
and the orbital angular frequency is given by \cite{c_Bambi_12}
\eq{Omega_varphi}{\Omega\equiv2\pi\nu_\varphi = \frac{-\partial_r g_{t\phi}\pm\sqrt{\left( \partial_rg_{t\varphi}\right)^2 - \partial_rg_{\varphi\varphi}\partial_rg_{tt}}}{\partial_rg_{\varphi\varphi}},}
where the positive (negative) sign in equation \Eref{Omega_varphi} refers to the co-rotating (counter-rotating) orbits with respect to the black hole spin. Equation 
\Eref{Omega_varphi} also defines the quantity $\nu_\varphi$ which is the frequency in which the particles move around the black hole in circular orbits.
Now the proper angular momentum ($\ell$) can be derived to be
\eq{curly_l}{\ell = -\frac{g_{t\varphi}+\Omega g_{\varphi\varphi}}{g_{tt}+\Omega g_{t\varphi}}.}

For finding the epicyclic frequencies, we first consider the perturbation to the radial $(r)$ and vertical $(\theta)$ 
coordinates so that \eq{per}{r(t)\approx r_0+\delta r(t)\text{,}\quad\theta(t)\approx \theta_0+\delta\theta(t),}
where the perturbations are considered to be $\delta r(t)\sim e^{i\Omega_rt}$ and $\delta\theta(t) \sim e^{i\Omega_\theta t}$, so as to have equations for harmonic oscillator of the form
\eq{radhar}{\frac{d^2\delta r}{dt^2} +\Omega_r^2\delta r = 0,}
\eq{anghar}{\frac{d^2\delta \theta}{dt^2} + \Omega_\theta^2\delta\theta = 0.}
Here  $r_0$ is the radius of the circular orbit and $\theta_0 = \pi/2$, is the angle at which the equatorial plane exists.
Now expanding the R.H.S. of equation \Eref{u_normalize_new} into second-order Taylor series along with the radial $(r)$ and vertical $(\theta)$ components,
replacing $r$ and $\theta$ from equation \Eref{per}, using equations \Eref{radhar} and \Eref{anghar}, 
and after some simple algebra we obtain \cite{c_Bambi_12,Ryan}

%Now the derivation of the epicyclic frequency requires the perturbation to the circular orbit slightly in both radial and vertical directions, such that
%\eq{}{r\approx r_0+\delta r e^{2\pi i\nu_rt}\text{,}\quad\theta\approx \theta_0+\delta\theta e^{2\pi i\nu_\theta t},}
%where $r_0$ is the radius of the circular orbit and $\theta_0 = \pi/2$ is the angle at which the equatorial plane exists. Substituting these in equation \Eref{u_normalize_new} and after some calculation ***WHAT CALCULATION*** the radial ($\nu_r$) and vertical ($\nu_\theta$) epicyclic frequencies can be found to be \cite{  Parthapratim_Pradhan,c_Bambi_12, Staykov}

\eq{nu_r}{\Omega_r^2 =\left(2\pi\nu_r\right)^2=- \frac{1}{2g_{rr}(u^t)^2}\frac{\partial^2\mathcal{V}_{eff}}{\partial r^2},}

\eq{nu_theta}{\Omega_\theta^2=\left(2\pi\nu_\theta\right)^2 =- \frac{1}{2g_{\theta\theta}(u^t)^2}\frac{\partial^2\mathcal{V}_{eff}}{\partial \theta^2}.}

The dependence of the frequencies on $B$ arises from various metric components. The explicit forms of the frequencies 
are huge and hence are not included in this work. Rather, we shall provide a numerical estimations of these frequencies. It should also be noted that these frequencies are observables and will be the key in estimating the most favored value of $B$ from observational data.

The behaviors of $\nu_r$ and $\nu_\theta$ are shown in Figures \ref{nu_r_for_a_1} and \ref{nu_theta_for_a_1} with a fixed spin parameter $a = 0.8$. From Figure \ref{four graphs} it can be seen that $\nu_r$ decreases, while $\nu_\theta$ 
and $\nu_\phi$ increase, with the increase of $|B|$, at a given $r$ 
(particularly away from the black hole). However, the peak of $\nu_\theta$ 
decreases with increasing $|B|$. Also $\nu_r$ vanishes at a larger radius with
a smaller peak with increasing $|B|$. It can be easily seen from equation \Eref{Omega_varphi} that the GR result, i.e. $\Omega \sim (r^{3/2}\pm a)^{-1}$, can be found by setting $B=0$.

\section{Conclusion}

The idea of modified GR is in the literature for sometime, but its indispensable
usefulness was not very clear. Although Starobinsky argued for $R^2$-gravity (a kind of
$f(R)$-gravity) to explain inflation \cite{1980PhLB...91...99S}, it was
not clear if all the gravity theories are the same. In last one decade or so, the authors
however showed that $R^2$-gravity could be useful to sort out problems lying with
neutron stars and white dwarfs \cite{PhysRevD.82.064033,Arapo_lu_2011,
Upasana15,Kalita18} as well. Nevertheless, none of these solutions is black hole
(vacuum) solution. In this work, we establish an asymptotically flat vacuum solution
of the axially symmetric field equation in a modified GR, more precisely $f(R)$-gravity.
The solution particularly describes the spacetime geometry around a rotating black hole,
i.e. the modified Kerr black hole solution, for the first time of this kind to the best
of our knowledge.

It shows that depending on the modified gravity
parameter, all the fundamental properties of the black hole change, e.g. the 
radii of black hole,
marginally stable and bound circular orbits increase. Therefore, based on the
observed size, e.g. by Event Horizon Telescope (EHT) image, the inference or
estimate of spin of black hole would be incorrect unless proper theory is used.
If indeed the gravity theory is based on an $f(R)$-gravity, the GR based inference of spin
of the black hole would actually underestimate it. This has many far reaching
astrophysical implications.

The solution also implies that the naked singularity, as formed at the Kerr parameter $a>1$,
need not necessarily produce in modified GR. This naturally has important implications to the
cosmic censorship hypothesis \cite{1969NCimR...1..252P,2002GReGr..34.1141P}.
Therefore, black holes, according to this gravity theory, can spin faster without forming
naked singularity depending on the modified gravity parameter.

\section*{Acknowledgement}
One of the authors (ARD) acknowledges the financial support from KVPY, DST, India.

\bibliographystyle{apsrev4-1} % We choose the "plain" reference style
\bibliography{refer}

%merlin.mbs apsrev4-1.bst 2010-07-25 4.21a (PWD, AO, DPC) hacked
%Control: key (0)
%Control: author (72) initials jnrlst
%Control: editor formatted (1) identically to author
%Control: production of article title (-1) disabled
%Control: page (0) single
%Control: year (1) truncated
%Control: production of eprint (0) enabled
\begin{thebibliography}{31}%
\makeatletter
\providecommand \@ifxundefined [1]{%
 \@ifx{#1\undefined}
}%
\providecommand \@ifnum [1]{%
 \ifnum #1\expandafter \@firstoftwo
 \else \expandafter \@secondoftwo
 \fi
}%
\providecommand \@ifx [1]{%
 \ifx #1\expandafter \@firstoftwo
 \else \expandafter \@secondoftwo
 \fi
}%
\providecommand \natexlab [1]{#1}%
\providecommand \enquote  [1]{``#1''}%
\providecommand \bibnamefont  [1]{#1}%
\providecommand \bibfnamefont [1]{#1}%
\providecommand \citenamefont [1]{#1}%
\providecommand \href@noop [0]{\@secondoftwo}%
\providecommand \href [0]{\begingroup \@sanitize@url \@href}%
\providecommand \@href[1]{\@@startlink{#1}\@@href}%
\providecommand \@@href[1]{\endgroup#1\@@endlink}%
\providecommand \@sanitize@url [0]{\catcode `\\12\catcode `\$12\catcode
  `\&12\catcode `\#12\catcode `\^12\catcode `\_12\catcode `\%12\relax}%
\providecommand \@@startlink[1]{}%
\providecommand \@@endlink[0]{}%
\providecommand \url  [0]{\begingroup\@sanitize@url \@url }%
\providecommand \@url [1]{\endgroup\@href {#1}{\urlprefix }}%
\providecommand \urlprefix  [0]{URL }%
\providecommand \Eprint [0]{\href }%
\providecommand \doibase [0]{http://dx.doi.org/}%
\providecommand \selectlanguage [0]{\@gobble}%
\providecommand \bibinfo  [0]{\@secondoftwo}%
\providecommand \bibfield  [0]{\@secondoftwo}%
\providecommand \translation [1]{[#1]}%
\providecommand \BibitemOpen [0]{}%
\providecommand \bibitemStop [0]{}%
\providecommand \bibitemNoStop [0]{.\EOS\space}%
\providecommand \EOS [0]{\spacefactor3000\relax}%
\providecommand \BibitemShut  [1]{\csname bibitem#1\endcsname}%
\let\auto@bib@innerbib\@empty
%</preamble>
\bibitem [{\citenamefont {et~al. [LIGO~Scientific}\ and\ \citenamefont
  {Collaborations]}(2016)}]{GW15}%
  \BibitemOpen
  \bibfield  {author} {\bibinfo {author} {\bibfnamefont {B.~P.~A.}\
  \bibnamefont {et~al. [LIGO~Scientific}}\ and\ \bibinfo {author}
  {\bibfnamefont {V.}~\bibnamefont {Collaborations]}},\ }\href {\doibase
  10.1103/physrevlett.116.061102} {\bibfield  {journal} {\bibinfo  {journal}
  {Physical Review Letters}\ }\textbf {\bibinfo {volume} {116}} (\bibinfo
  {year} {2016}),\ 10.1103/physrevlett.116.061102}\BibitemShut {NoStop}%
\bibitem [{\citenamefont {Das}\ and\ \citenamefont
  {Mukhopadhyay}(2015)}]{Upasana15}%
  \BibitemOpen
  \bibfield  {author} {\bibinfo {author} {\bibfnamefont {U.}~\bibnamefont
  {Das}}\ and\ \bibinfo {author} {\bibfnamefont {B.}~\bibnamefont
  {Mukhopadhyay}},\ }\href {\doibase 10.1088/1475-7516/2015/05/045} {\bibfield
  {journal} {\bibinfo  {journal} {Journal of Cosmology and Astroparticle
  Physics}\ }\textbf {\bibinfo {volume} {2015}},\ \bibinfo {pages} {045}
  (\bibinfo {year} {2015})}\BibitemShut {NoStop}%
\bibitem [{\citenamefont {Kalita}\ and\ \citenamefont
  {Mukhopadhyay}(2018)}]{Kalita18}%
  \BibitemOpen
  \bibfield  {author} {\bibinfo {author} {\bibfnamefont {S.}~\bibnamefont
  {Kalita}}\ and\ \bibinfo {author} {\bibfnamefont {B.}~\bibnamefont
  {Mukhopadhyay}},\ }\href {\doibase 10.1088/1475-7516/2018/09/007} {\bibfield
  {journal} {\bibinfo  {journal} {Journal of Cosmology and Astroparticle
  Physics}\ }\textbf {\bibinfo {volume} {2018}},\ \bibinfo {pages} {007–007}
  (\bibinfo {year} {2018})}\BibitemShut {NoStop}%
\bibitem [{\citenamefont {Kalita}\ and\ \citenamefont
  {Mukhopadhyay}(2019)}]{Kalita_Bani}%
  \BibitemOpen
  \bibfield  {author} {\bibinfo {author} {\bibfnamefont {S.}~\bibnamefont
  {Kalita}}\ and\ \bibinfo {author} {\bibfnamefont {B.}~\bibnamefont
  {Mukhopadhyay}},\ }\href {\doibase 10.1140/epjc/s10052-019-7396-x} {\bibfield
   {journal} {\bibinfo  {journal} {The European Physical Journal C}\ }\textbf
  {\bibinfo {volume} {79}} (\bibinfo {year} {2019}),\
  10.1140/epjc/s10052-019-7396-x}\BibitemShut {NoStop}%
\bibitem [{\citenamefont {Nojiri}\ and\ \citenamefont
  {Odintsov}(2013)}]{Nojiri:2013su}%
  \BibitemOpen
  \bibfield  {author} {\bibinfo {author} {\bibfnamefont {S.}~\bibnamefont
  {Nojiri}}\ and\ \bibinfo {author} {\bibfnamefont {S.~D.}\ \bibnamefont
  {Odintsov}},\ }\href {\doibase 10.1088/0264-9381/30/12/125003} {\bibfield
  {journal} {\bibinfo  {journal} {Classical and Quantum Gravity}\ }\textbf
  {\bibinfo {volume} {30}},\ \bibinfo {pages} {125003} (\bibinfo {year}
  {2013})}\BibitemShut {NoStop}%
\bibitem [{\citenamefont {Nojiri}\ and\ \citenamefont
  {Odintsov}(2017)}]{Nojiri:2017kex}%
  \BibitemOpen
  \bibfield  {author} {\bibinfo {author} {\bibfnamefont {S.}~\bibnamefont
  {Nojiri}}\ and\ \bibinfo {author} {\bibfnamefont {S.}~\bibnamefont
  {Odintsov}},\ }\href {\doibase 10.1103/physrevd.96.104008} {\bibfield
  {journal} {\bibinfo  {journal} {Physical Review D}\ }\textbf {\bibinfo
  {volume} {96}} (\bibinfo {year} {2017}),\
  10.1103/physrevd.96.104008}\BibitemShut {NoStop}%
\bibitem [{\citenamefont {Nojiri}\ \emph {et~al.}(2021)\citenamefont {Nojiri},
  \citenamefont {Odintsov},\ and\ \citenamefont {Faraoni}}]{Nojiri:2020blr}%
  \BibitemOpen
  \bibfield  {author} {\bibinfo {author} {\bibfnamefont {S.}~\bibnamefont
  {Nojiri}}, \bibinfo {author} {\bibfnamefont {S.~D.}\ \bibnamefont
  {Odintsov}}, \ and\ \bibinfo {author} {\bibfnamefont {V.}~\bibnamefont
  {Faraoni}},\ }\href {\doibase 10.1103/physrevd.103.044055} {\bibfield
  {journal} {\bibinfo  {journal} {Physical Review D}\ }\textbf {\bibinfo
  {volume} {103}} (\bibinfo {year} {2021}),\
  10.1103/physrevd.103.044055}\BibitemShut {NoStop}%
\bibitem [{\citenamefont {Newman}\ and\ \citenamefont
  {Janis}(1965)}]{NJA_Orig}%
  \BibitemOpen
  \bibfield  {author} {\bibinfo {author} {\bibfnamefont {E.~T.}\ \bibnamefont
  {Newman}}\ and\ \bibinfo {author} {\bibfnamefont {A.~I.}\ \bibnamefont
  {Janis}},\ }\href {\doibase 10.1063/1.1704350} {\bibfield  {journal}
  {\bibinfo  {journal} {Journal of Mathematical Physics}\ }\textbf {\bibinfo
  {volume} {6}},\ \bibinfo {pages} {915} (\bibinfo {year} {1965})},\ \Eprint
  {http://arxiv.org/abs/https://doi.org/10.1063/1.1704350}
  {https://doi.org/10.1063/1.1704350} \BibitemShut {NoStop}%
\bibitem [{\citenamefont {Misner}\ \emph {et~al.}(1973)\citenamefont {Misner},
  \citenamefont {Thorne},\ and\ \citenamefont
  {Wheeler}}]{Misner_Thorne_Wheeler_73}%
  \BibitemOpen
  \bibfield  {author} {\bibinfo {author} {\bibfnamefont {C.~W.}\ \bibnamefont
  {Misner}}, \bibinfo {author} {\bibfnamefont {K.~S.}\ \bibnamefont {Thorne}},
  \ and\ \bibinfo {author} {\bibfnamefont {J.~A.}\ \bibnamefont {Wheeler}},\
  }\href@noop {} {\emph {\bibinfo {title} {{Gravitation}}}}\ (\bibinfo
  {publisher} {W. H. Freeman},\ \bibinfo {address} {San Francisco},\ \bibinfo
  {year} {1973})\BibitemShut {NoStop}%
\bibitem [{\citenamefont {De~Felice}\ and\ \citenamefont
  {Tsujikawa}(2010)}]{Felice_Tsujikawa_10}%
  \BibitemOpen
  \bibfield  {author} {\bibinfo {author} {\bibfnamefont {A.}~\bibnamefont
  {De~Felice}}\ and\ \bibinfo {author} {\bibfnamefont {S.}~\bibnamefont
  {Tsujikawa}},\ }\href {\doibase 10.12942/lrr-2010-3} {\bibfield  {journal}
  {\bibinfo  {journal} {Living Reviews in Relativity}\ }\textbf {\bibinfo
  {volume} {13}} (\bibinfo {year} {2010}),\ 10.12942/lrr-2010-3}\BibitemShut
  {NoStop}%
\bibitem [{\citenamefont {Nojiri}\ \emph {et~al.}(2017)\citenamefont {Nojiri},
  \citenamefont {Odintsov},\ and\ \citenamefont
  {Oikonomou}}]{Nojiri_Odintsov_Oikonomou_17}%
  \BibitemOpen
  \bibfield  {author} {\bibinfo {author} {\bibfnamefont {S.}~\bibnamefont
  {Nojiri}}, \bibinfo {author} {\bibfnamefont {S.}~\bibnamefont {Odintsov}}, \
  and\ \bibinfo {author} {\bibfnamefont {V.}~\bibnamefont {Oikonomou}},\ }\href
  {\doibase https://doi.org/10.1016/j.physrep.2017.06.001} {\bibfield
  {journal} {\bibinfo  {journal} {Physics Reports}\ }\textbf {\bibinfo {volume}
  {692}},\ \bibinfo {pages} {1} (\bibinfo {year} {2017})},\ \bibinfo {note}
  {modified Gravity Theories on a Nutshell: Inflation, Bounce and Late-time
  Evolution}\BibitemShut {NoStop}%
\bibitem [{\citenamefont {Nojiri}\ and\ \citenamefont
  {Odintsov}(2011)}]{Nojiri:2010wj}%
  \BibitemOpen
  \bibfield  {author} {\bibinfo {author} {\bibfnamefont {S.}~\bibnamefont
  {Nojiri}}\ and\ \bibinfo {author} {\bibfnamefont {S.~D.}\ \bibnamefont
  {Odintsov}},\ }\href {\doibase 10.1016/j.physrep.2011.04.001} {\bibfield
  {journal} {\bibinfo  {journal} {Physics Reports}\ }\textbf {\bibinfo {volume}
  {505}},\ \bibinfo {pages} {59} (\bibinfo {year} {2011})}\BibitemShut
  {NoStop}%
\bibitem [{\citenamefont {Multam\"aki}\ and\ \citenamefont
  {Vilja}(2006)}]{Mut_Vij_06}%
  \BibitemOpen
  \bibfield  {author} {\bibinfo {author} {\bibfnamefont {T.}~\bibnamefont
  {Multam\"aki}}\ and\ \bibinfo {author} {\bibfnamefont {I.}~\bibnamefont
  {Vilja}},\ }\href {\doibase 10.1103/PhysRevD.74.064022} {\bibfield  {journal}
  {\bibinfo  {journal} {Physical Review D}\ }\textbf {\bibinfo {volume} {74}},\
  \bibinfo {pages} {064022} (\bibinfo {year} {2006})}\BibitemShut {NoStop}%
\bibitem [{\citenamefont {Weinberg}(1972)}]{Weinberg_1972}%
  \BibitemOpen
  \bibfield  {author} {\bibinfo {author} {\bibfnamefont {S.}~\bibnamefont
  {Weinberg}},\ }\href@noop {} {\emph {\bibinfo {title} {{Gravitation and
  Cosmology}: {Principles and Applications of the General Theory of
  Relativity}}}}\ (\bibinfo  {publisher} {John Wiley and Sons},\ \bibinfo
  {address} {New York},\ \bibinfo {year} {1972})\BibitemShut {NoStop}%
\bibitem [{\citenamefont {Drake}\ and\ \citenamefont {Szekeres}(2000)}]{NJA_1}%
  \BibitemOpen
  \bibfield  {author} {\bibinfo {author} {\bibfnamefont {S.~P.}\ \bibnamefont
  {Drake}}\ and\ \bibinfo {author} {\bibfnamefont {P.}~\bibnamefont
  {Szekeres}},\ }\href {\doibase 10.1023/a:1001920232180} {\bibfield  {journal}
  {\bibinfo  {journal} {General Relativity and Gravitation}\ }\textbf {\bibinfo
  {volume} {32}},\ \bibinfo {pages} {445–457} (\bibinfo {year}
  {2000})}\BibitemShut {NoStop}%
\bibitem [{\citenamefont {Brauer}\ \emph {et~al.}(2014)\citenamefont {Brauer},
  \citenamefont {Camargo},\ and\ \citenamefont {Socolovsky}}]{NJA_Revisited}%
  \BibitemOpen
  \bibfield  {author} {\bibinfo {author} {\bibfnamefont {O.}~\bibnamefont
  {Brauer}}, \bibinfo {author} {\bibfnamefont {H.~A.}\ \bibnamefont {Camargo}},
  \ and\ \bibinfo {author} {\bibfnamefont {M.}~\bibnamefont {Socolovsky}},\
  }\href {\doibase 10.1007/s10773-014-2225-3} {\bibfield  {journal} {\bibinfo
  {journal} {International Journal of Theoretical Physics}\ }\textbf {\bibinfo
  {volume} {54}},\ \bibinfo {pages} {302–314} (\bibinfo {year}
  {2014})}\BibitemShut {NoStop}%
\bibitem [{\citenamefont {Azreg-A\"{\i}nou}(2014)}]{azeg}%
  \BibitemOpen
  \bibfield  {author} {\bibinfo {author} {\bibfnamefont {M.}~\bibnamefont
  {Azreg-A\"{\i}nou}},\ }\href {\doibase 10.1103/PhysRevD.90.064041} {\bibfield
   {journal} {\bibinfo  {journal} {Physical Review D}\ }\textbf {\bibinfo
  {volume} {90}},\ \bibinfo {pages} {064041} (\bibinfo {year}
  {2014})}\BibitemShut {NoStop}%
\bibitem [{\citenamefont {Azreg-A{\"i}nou}(2014)}]{azeg2}%
  \BibitemOpen
  \bibfield  {author} {\bibinfo {author} {\bibfnamefont {M.}~\bibnamefont
  {Azreg-A{\"i}nou}},\ }\href {\doibase 10.1140/epjc/s10052-014-2865-8}
  {\bibfield  {journal} {\bibinfo  {journal} {The European Physical Journal C}\
  }\textbf {\bibinfo {volume} {74}},\ \bibinfo {pages} {2865} (\bibinfo {year}
  {2014})}\BibitemShut {NoStop}%
\bibitem [{\citenamefont {Azreg-Aïnou}(2014)}]{azeg3}%
  \BibitemOpen
  \bibfield  {author} {\bibinfo {author} {\bibfnamefont {M.}~\bibnamefont
  {Azreg-Aïnou}},\ }\href {\doibase 10.1016/j.physletb.2014.01.041} {\bibfield
   {journal} {\bibinfo  {journal} {Physics Letters B}\ }\textbf {\bibinfo
  {volume} {730}},\ \bibinfo {pages} {95} (\bibinfo {year} {2014})}\BibitemShut
  {NoStop}%
\bibitem [{\citenamefont {Erbin}(2017)}]{Source_and_singularity}%
  \BibitemOpen
  \bibfield  {author} {\bibinfo {author} {\bibfnamefont {H.}~\bibnamefont
  {Erbin}},\ }\href {\doibase 10.3390/universe3010019} {\bibfield  {journal}
  {\bibinfo  {journal} {Universe}\ }\textbf {\bibinfo {volume} {3}},\ \bibinfo
  {pages} {19} (\bibinfo {year} {2017})}\BibitemShut {NoStop}%
\bibitem [{\citenamefont {Aliev}\ and\ \citenamefont {G\"umr\"uk\ifmmode
  \mbox{\c{c}}\else \c{c}\fi{}\"uo\ifmmode~\breve{g}\else
  \u{g}\fi{}lu}(2005)}]{Charged_BH_Brane}%
  \BibitemOpen
  \bibfield  {author} {\bibinfo {author} {\bibfnamefont {A.~N.}\ \bibnamefont
  {Aliev}}\ and\ \bibinfo {author} {\bibfnamefont {A.~E.}\ \bibnamefont
  {G\"umr\"uk\ifmmode \mbox{\c{c}}\else \c{c}\fi{}\"uo\ifmmode~\breve{g}\else
  \u{g}\fi{}lu}},\ }\href {\doibase 10.1103/PhysRevD.71.104027} {\bibfield
  {journal} {\bibinfo  {journal} {Physical Review D}\ }\textbf {\bibinfo
  {volume} {71}},\ \bibinfo {pages} {104027} (\bibinfo {year}
  {2005})}\BibitemShut {NoStop}%
\bibitem [{\citenamefont {Dadhich}\ \emph {et~al.}(2000)\citenamefont
  {Dadhich}, \citenamefont {Maartens}, \citenamefont {Papadopoulos},\ and\
  \citenamefont {Rezania}}]{BH_Brane}%
  \BibitemOpen
  \bibfield  {author} {\bibinfo {author} {\bibfnamefont {N.}~\bibnamefont
  {Dadhich}}, \bibinfo {author} {\bibfnamefont {R.}~\bibnamefont {Maartens}},
  \bibinfo {author} {\bibfnamefont {P.}~\bibnamefont {Papadopoulos}}, \ and\
  \bibinfo {author} {\bibfnamefont {V.}~\bibnamefont {Rezania}},\ }\href
  {\doibase 10.1016/s0370-2693(00)00798-x} {\bibfield  {journal} {\bibinfo
  {journal} {Physics Letters B}\ }\textbf {\bibinfo {volume} {487}},\ \bibinfo
  {pages} {1–6} (\bibinfo {year} {2000})}\BibitemShut {NoStop}%
\bibitem [{\citenamefont {{Hartle}}\ and\ \citenamefont
  {{Traschen}}(2005)}]{Hartle_Gravity_book}%
  \BibitemOpen
  \bibfield  {author} {\bibinfo {author} {\bibfnamefont {J.~B.}\ \bibnamefont
  {{Hartle}}}\ and\ \bibinfo {author} {\bibfnamefont {J.}~\bibnamefont
  {{Traschen}}},\ }\href {\doibase 10.1063/1.2405550} {\bibfield  {journal}
  {\bibinfo  {journal} {Physics Today}\ }\textbf {\bibinfo {volume} {58}},\
  \bibinfo {pages} {52} (\bibinfo {year} {2005})}\BibitemShut {NoStop}%
\bibitem [{\citenamefont {Shapiro}\ and\ \citenamefont
  {Teukolsky}(1983)}]{STbook}%
  \BibitemOpen
  \bibfield  {author} {\bibinfo {author} {\bibfnamefont {S.}~\bibnamefont
  {Shapiro}}\ and\ \bibinfo {author} {\bibfnamefont {S.}~\bibnamefont
  {Teukolsky}},\ }\href@noop {} {\emph {\bibinfo {title} {{Black Holes, White
  Dwarfs, and Neutron Stars: The Physics of Compact Objects}}}}\ (\bibinfo
  {publisher} {John Wiley and Sons},\ \bibinfo {address} {Weinheim},\ \bibinfo
  {year} {1983})\BibitemShut {NoStop}%
\bibitem [{\citenamefont {Bambi}(2012)}]{c_Bambi_12}%
  \BibitemOpen
  \bibfield  {author} {\bibinfo {author} {\bibfnamefont {C.}~\bibnamefont
  {Bambi}},\ }\href {\doibase 10.1088/1475-7516/2012/09/014} {\bibfield
  {journal} {\bibinfo  {journal} {Journal of Cosmology and Astroparticle
  Physics}\ }\textbf {\bibinfo {volume} {2012}},\ \bibinfo {pages} {014–014}
  (\bibinfo {year} {2012})}\BibitemShut {NoStop}%
\bibitem [{\citenamefont {Ryan}(1995)}]{Ryan}%
  \BibitemOpen
  \bibfield  {author} {\bibinfo {author} {\bibfnamefont {F.~D.}\ \bibnamefont
  {Ryan}},\ }\href {\doibase 10.1103/PhysRevD.52.5707} {\bibfield  {journal}
  {\bibinfo  {journal} {Physical Review D}\ }\textbf {\bibinfo {volume} {52}},\
  \bibinfo {pages} {5707} (\bibinfo {year} {1995})}\BibitemShut {NoStop}%
\bibitem [{\citenamefont {{Starobinsky}}(1980)}]{1980PhLB...91...99S}%
  \BibitemOpen
  \bibfield  {author} {\bibinfo {author} {\bibfnamefont {A.~A.}\ \bibnamefont
  {{Starobinsky}}},\ }\href {\doibase 10.1016/0370-2693(80)90670-X} {\bibfield
  {journal} {\bibinfo  {journal} {Physics Letters B}\ }\textbf {\bibinfo
  {volume} {91}},\ \bibinfo {pages} {99} (\bibinfo {year} {1980})}\BibitemShut
  {NoStop}%
\bibitem [{\citenamefont {Cooney}\ \emph {et~al.}(2010)\citenamefont {Cooney},
  \citenamefont {DeDeo},\ and\ \citenamefont {Psaltis}}]{PhysRevD.82.064033}%
  \BibitemOpen
  \bibfield  {author} {\bibinfo {author} {\bibfnamefont {A.}~\bibnamefont
  {Cooney}}, \bibinfo {author} {\bibfnamefont {S.}~\bibnamefont {DeDeo}}, \
  and\ \bibinfo {author} {\bibfnamefont {D.}~\bibnamefont {Psaltis}},\ }\href
  {\doibase 10.1103/PhysRevD.82.064033} {\bibfield  {journal} {\bibinfo
  {journal} {Physical Review D}\ }\textbf {\bibinfo {volume} {82}},\ \bibinfo
  {pages} {064033} (\bibinfo {year} {2010})}\BibitemShut {NoStop}%
\bibitem [{\citenamefont {Arapo{\u{g}}lu}\ \emph {et~al.}(2011)\citenamefont
  {Arapo{\u{g}}lu}, \citenamefont {Deliduman},\ and\ \citenamefont
  {Ek{\c{s}}i}}]{Arapo_lu_2011}%
  \BibitemOpen
  \bibfield  {author} {\bibinfo {author} {\bibfnamefont {S.}~\bibnamefont
  {Arapo{\u{g}}lu}}, \bibinfo {author} {\bibfnamefont {C.}~\bibnamefont
  {Deliduman}}, \ and\ \bibinfo {author} {\bibfnamefont {K.~Y.}\ \bibnamefont
  {Ek{\c{s}}i}},\ }\href {\doibase 10.1088/1475-7516/2011/07/020} {\bibfield
  {journal} {\bibinfo  {journal} {Journal of Cosmology and Astroparticle
  Physics}\ }\textbf {\bibinfo {volume} {2011}},\ \bibinfo {pages} {020}
  (\bibinfo {year} {2011})}\BibitemShut {NoStop}%
\bibitem [{\citenamefont {{Penrose}}(1969)}]{1969NCimR...1..252P}%
  \BibitemOpen
  \bibfield  {author} {\bibinfo {author} {\bibfnamefont {R.}~\bibnamefont
  {{Penrose}}},\ }\href@noop {} {\bibfield  {journal} {\bibinfo  {journal}
  {Nuovo Cimento Rivista Serie}\ }\textbf {\bibinfo {volume} {1}},\ \bibinfo
  {pages} {252} (\bibinfo {year} {1969})}\BibitemShut {NoStop}%
\bibitem [{\citenamefont {{Penrose}}(2002)}]{2002GReGr..34.1141P}%
  \BibitemOpen
  \bibfield  {author} {\bibinfo {author} {\bibfnamefont {R.}~\bibnamefont
  {{Penrose}}},\ }\href {\doibase 10.1023/A:1016578408204} {\bibfield
  {journal} {\bibinfo  {journal} {General Relativity and Gravitation}\ }\textbf
  {\bibinfo {volume} {7}},\ \bibinfo {pages} {1141} (\bibinfo {year}
  {2002})}\BibitemShut {NoStop}%
\end{thebibliography}%

\end{document}